\documentclass{ws-ijmpa}
\usepackage[super,compress]{cite}
\usepackage{graphicx}
\usepackage{color}
\begin{document}
\markboth{H. Fukaya}{Understanding the index theorems with massive fermions}

%
%

\title{Understanding the index theorems with massive fermions
}

\author{Hidenori Fukaya}

\address{Department of Physics, Osaka University, Machikane-yama-cho 1-1, Toyonaka 560-0043, Japan}

\maketitle

\begin{history}
\end{history}

\begin{abstract}
  The index theorems relate the gauge field and metric on a manifold
  to the solution of the Dirac equation on it.
  In the standard approach, the Dirac operator must be massless
  in order to make the chirality operator well-defined.
  In physics, however, the index theorem appears as a consequence of
  chiral anomaly, which is an explicit breaking of the symmetry.
  It is then natural to ask if we can understand the index theorems
  in a massive fermion system which does not have chiral symmetry.
  In this review\footnote{
    This article is based on lectures at YITP, Kyoto university, in December 2020.  
  }, we discuss how to reformulate the chiral anomaly and index theorems
  with massive Dirac operators, where we find nontrivial mathematical relations
  between massless and massive fermions.
  A special focus is placed on the Atiyah-Patodi-Singer index,
  whose original formulation requires a physicist-unfriendly boundary condition,
  while the corresponding massive domain-wall fermion reformulation does not.
  The massive formulation provides a natural understanding of the
  anomaly inflow between the bulk and edge in particle and condensed matter physics.
\end{abstract}

\ccode{PACS numbers:}


\section{Introduction}	
\label{sec:intro}
The Atiyah-Singer (AS) index theorem \cite{Atiyah:1963zz,Atiyah:1968mp}
on a manifold without boundary
is well understood and appreciated in physics.
The theorem relates the number of solutions of a Dirac equation
having a definite chirality to the topological invariant
given by the gauge field and metric on it.
In particle physics, it gives, for example, an essential understanding
of a non-perturbative tunneling effect in the vacuum of
quantum chromodynamics (QCD) through instantons.

The Atiyah-Patodi-Singer (APS) index theorem \cite{Atiyah:1975jf,Atiyah:1976jg, Atiyah:1980jh},
which is an extension of the AS theorem to a manifold with boundaries
is, however, not very physicist-friendly.
In order to keep the chirality well-defined, 
a nonlocal boundary condition is imposed by hand,
which is known as the APS boundary condition.
In relativistic physics, any condition on fields must be locally given,
otherwise, the causality may be lost as
information propagates faster than the speed of light.

For an intuitive understanding, let us consider a massless free
fermionic particle reflecting at a boundary,
which is a flat Euclidean compact manifold $Y$.
To conserve the energy and momentum in the horizontal
directions, the particle must flip the momentum
in the normal direction to $Y$.
On the other hand, the spin of the particle should not change
as $Y$ has a rotational symmetry with the axis
perpendicular to $Y$.
In this case, the helicity or the spin in the moving direction
of the particle
flips and the chiral symmetry is lost.
This simple example tells us that it is natural
to lose chiral symmetry on a manifold with boundary
when a local and physically sensible boundary condition is imposed.
Otherwise, a non-local and physically implausible boundary condition is needed.

It is interesting to note that
the index theorems appear in physics as a consequence
of quantum anomaly\cite{Adler:1969gk,Bell:1969ts} of the chiral symmetry.
The index itself counts the mismatch of
the left-handed and right-handed modes of fermions.
We may say that the index is defined with a symmetry,
which is explicitly broken by quantum anomaly.
Then it is natural to ask if we can reformulate
the index theorems without chiral symmetry from the beginning,
or equivalently, if we can reformulate them with massive fermions.
In this review, we would like to show that the answer is ``Yes, we can.''

The key to describe the index theorems without chiral symmetry
is the so-called domain-wall fermion \cite{Jackiw:1975fn, Callan:1984sa, Kaplan:1992bt}. 
The domain-wall fermion is a massive fermion
with a mass term flipping its sign at some codimension-one submanifold.
Inside bulk, or regions far from the wall, the fermion
is gapped or massive, while a massless edge mode appears
localized at the domain-wall.
The domain-wall fermion provides a good model for  a topological insulator
surrounded by a normal insulator.
We can regard the negative mass (compared to that of regulator)
regions as a topological matter
and the positive ones as in the normal phase.

In Ref.~\citen{Fukaya:2017tsq}, we perturbatively showed
that the $\eta$ invariant of a massive domain-wall fermion
Dirac operator on a flat four-dimensional Euclidean manifold 
coincides with the APS index on a ``half'' of
the same manifold or the negative mass region
with the APS boundary condition assigned to the location of the domain-wall.
After Ref.~\citen{Fukaya:2017tsq}, three mathematicians joined our collaboration
and we achieved a mathematical proof \cite{Fukaya:2019qlf} that
the equality is mathematically justified:
for any APS index on
a general Riemannian manifold with boundary,
there exists a domain-wall Dirac operator on an extended manifold
attaching ``outside'' of the original one, whose $\eta$ invariant
is equal to the original APS index.
Our massive formulation is so physicist-friendly that
application to lattice gauge theory is straightforward \cite{Fukaya:2019myi}
(see Ref.~\citen{Onogi:2021slv} for a relation to the Berry phase).
Recently we extended our work to the mod-two APS index\cite{Fukaya:2020tjk},
which is defined on odd-dimensional manifolds.

Our work has a tight connection to the anomaly inflow\cite{Callan:1984sa, Witten:2015aba},
or anomaly matching \cite{tHooft:1979rat} between bulk and edge \cite{Hatsugai}
fermions, which attracts a significant attention 
in particle physics 
\cite{Kurkov:2018pjw,Tachikawa:2018njr,Vassilevich:2018aqu,Garcia-Etxebarria:2018ajm,Yonekura:2019vyz,Hsieh:2020jpj,Hamada:2020mug}
and condensed matter physics
\cite{Gromov:2015fda,Metlitski:2015yqa,Seiberg:2016rsg,Tachikawa:2016xvs,Freed:2016rqq,Yu:2017uqt,Hasebe:2016tjg,Yonekura:2018ufj,Yao:2019ggu}.
As will be shown below, the roles of bulk and edge modes
are manifest in our massive reformulation
and we can intuitively understand how their anomaly is canceled,
in contrast to the case with the APS boundary condition,
which does not allow any edge-localized modes to exist.

The rest of this article is organized as follows.
We start in Sec.~\ref{sec:U1anomaly}
with the standard axial $U(1)$ anomaly
employing the Pauli-Villars regularization.
We will see that the anomaly comes from the mass term of the Pauli-Villars field.
In Sec.~\ref{sec:AS}, we try to reformulate the
Atiyah-Singer index on a closed manifold with massive Dirac operator.
Then we review the original work of APS in Sec.~\ref{sec:APSreview}
and explain why it is physicist-unfriendly.
In Sec.~\ref{sec:APSDW}, we reformulate the APS index
with the domain-wall fermion Dirac operator.
The application to the lattice gauge theory (Sec.~\ref{sec:APSlattice})
and mod-two index on odd-dimensional manifold (Sec.~\ref{sec:APSmod2})
are briefly reviewed.
Summary and discussion are given in Sec.~\ref{sec:summary}.

\section{Perturbative computation of axial $U(1)$ anomaly}
\label{sec:U1anomaly}

As a warm-up, let us discuss the axial $U(1)$ anomaly\cite{Adler:1969gk,Bell:1969ts}. 
In the textbooks, this anomaly is beautifully obtained
by the Fujikawa method with the heat-kernel regularization.
Here we revisit this computation with the Pauli-Villars(PV) regularization\cite{Fujikawa:2004cx},
which is slightly tedious but makes it clear that 
the anomaly is an explicit symmetry breaking of the theory
originating from the mass term.

We start with the massless Dirac fermion action
\begin{align}
S = \int_X d^4x\; \bar{\psi}D\psi(x),
\end{align}
on a four-dimensional Euclidean flat space $X$,
where $\psi$ and $\bar{\psi}$ are four-component spinors
on which the Dirac operator $D=\gamma^\mu(\partial_\mu + iA_\mu)$ operates.
The gamma matrices satisfy $\{\gamma_\mu,\gamma_\nu\}=2\delta_{\mu\nu}$
and we take the $SU(N)$ gauge field $A_\mu= \sum_a A_\mu^a T^a$
with Hermitian generators $T^a$.

With the chirality operator $\gamma_5 = -\gamma_1\gamma_2\gamma_3\gamma_4$,
the action is invariant under the axial $U(1)$ rotation:
\begin{align}
  \psi&\to\exp(i\alpha \gamma_5)\psi,\;\;\;
\bar{\psi}\to\bar{\psi}\exp(i\alpha \gamma_5),
\end{align}
since $e^{i\alpha \gamma_5} D e^{i\alpha \gamma_5} = D$.
For $x$-dependent angle $\alpha(x)$ we have
\begin{align}
 e^{i\alpha(x) \gamma_5} D e^{i\alpha(x) \gamma_5} = D + i\gamma^\mu\gamma_5\partial_\mu \alpha(x).
\end{align}
Assuming $\alpha(x)\to 0$ at $|x|\to \infty$  we obtain by a partial integration that
\begin{align}
S &= \int_X d^4x\; \bar{\psi}D\psi(x)
-i\int_X d^4x\;\alpha(x)\partial_\mu\left[\bar{\psi}\gamma^\mu\gamma_5\psi(x)\right],
\end{align}
from which we may classically conclude the conservation of the axial current
$J^\mu_5(x)=\bar{\psi}\gamma^\mu\gamma_5\psi(x)$.

However, in quantum field theory, we have to take the path-integral measure into account:
\begin{align}
Z
&= \int [D\bar{\psi}D\psi]J e^{-S[\bar{\psi},\psi]+\int d^4x \alpha(x)\partial_\mu J_5^\mu(x)},
\end{align}
where $J$ is the Jacobian of the measure.
Fujikawa showed that $J\neq 1$ and $\partial_\mu J_5^\mu(x)$ does not conserve.
This is the standard derivation of the axial $U(1)$ anomaly.

Here we do not directly compute $J$ but instead introduce
a bosonic spinor field $\phi(x)$ and add its action\footnote{
  In general we need more PV fields to fully regularize the theory.
  But for the computation of anomaly one bosonic spinor is enough.
} 
\begin{align}
  S = \int_X d^4x\; \bar{\psi}D\psi(x) + \int_X d^4x\; \bar{\phi}(D+M)\phi(x),
\end{align}
where $M$ is a cut-off scale mass. Then the path-integral becomes
\begin{align}
Z=\int [D\bar{\psi}D\psi]\int [D\bar{\phi}D\phi]e^{-S} = \frac{\det D}{\det (D+M)}.
\end{align}

Let us perform the axial $U(1)$ rotation on $\psi$ and $\phi$:
\begin{align}
Z&= \int [D\bar{\psi}D\psi]J\int [D\bar{\phi}D\phi]J^{-1}
e^{-S[\bar{\psi},\psi,\bar{\phi},\phi]+\int d^4x \alpha(x)\left[\partial_\mu J_{5,PV}^\mu(x)
+2i\bar{\phi}M\gamma_5\phi(x)\right]},
\end{align}
where $J_{5,{\rm PV}}^\mu(x) = \bar{\psi}\gamma^\mu\gamma_5\psi(x)+\bar{\phi}\gamma^\mu\gamma_5\phi(x)$
is the PV-regularized axial current.
We can see that the Jacobian of $\psi$ field is precisely canceled
by that of $\phi$. Instead, we have a change in the mass term $2i\bar{\phi}M\gamma_5\phi(x)$.

Now we have a relation
\begin{align}
  \label{eq:PVanomaly}
  \partial_\mu\left\langle J_{5,{\rm PV}}^\mu(x)\right\rangle &=
  2i\left\langle\bar{\phi}M\gamma_5\phi(x)\right\rangle = 2M{\rm tr}\left[\gamma_5 \frac{1}{D+M}\right](x,x),
\end{align}
where $\langle O\rangle$ denotes the functional average of an operator $O$.
Is this the axial anomaly? The answer is definitely ``yes''.
It is a good exercise in the large $M$ limit to reproduce the
anomaly. Using the same properties of $f(x)=1/(1+x)$ with the standard heat-kernel $f(x)=e^{-x}$,
such as $f(0)=1$, $f(\infty)=0$ and $n$-th derivative $f^{(n)}(x)x |_{x=0} =0$,
it is straightforward to confirm the right hand side (RHS) of Eq.~(\ref{eq:PVanomaly}) becomes
\begin{align}
=\frac{2}{32\pi^2}\varepsilon^{\mu\nu\rho\sigma}
{\rm tr}F_{\mu\nu}F_{\rho\sigma}(x),
\end{align}
where $F_{\mu\nu}=\partial_\mu A_\nu-\partial_\nu A_\mu+i[A_\mu,A_\nu]$
is the field strength.
It is thus obvious that the axial $U(1)$ anomaly is an explicit breaking
as it originates from the mass term of the Pauli-Villars field.

Next, let us discuss the integral of the anomaly
over a compact and closed manifold $X$\footnote{For example, $X$ is a four-dimensional torus $X=T^4$.},
\begin{align}
  \int_X d^4x M{\rm tr}\left[\gamma_5 \frac{1}{D+M}\right](x,x)=\frac{1}{32\pi^2}\int_X d^4x \varepsilon^{\mu\nu\rho\sigma}
       {\rm tr}F_{\mu\nu}F_{\rho\sigma}(x),
\end{align}
where the RHS is known to give an integer called the winding number.
The left hand side (LHS) can be expressed by a spectral decomposition,
\begin{align}
  \label{eq:gamma5trace}
\int d^4x M{\rm tr}\left[\gamma_5 \frac{1}{D+M}\right](x,x)
&= {\rm Tr}\left[\gamma_5\frac{1}{1-D^2/M^2}\right]
=\sum_{\lambda}\langle \lambda| \gamma_5 |\lambda\rangle \frac{1}{1+\lambda^2/M^2}
\end{align}
where ${\rm Tr}=\int_X d^4x\;{\rm tr}$ and we have inserted
the eigenmode complete set of the Dirac operator,
$D |\lambda\rangle =i \lambda |\lambda\rangle$.
Every nonzero eigenmode $\lambda\neq 0$ makes a pair with
the one with the opposite sign since
\begin{equation}
D \gamma_5|\lambda\rangle =-\gamma_5 D |\lambda\rangle = -i\lambda \gamma_5|\lambda\rangle \propto |-\lambda\rangle
\end{equation}
and therefore, $\langle \lambda |\gamma_5 |\lambda\rangle  \propto \langle \lambda |-\lambda\rangle = 0$.
We now obtain the Atiyah-Singer index theorem,
\begin{align}
n_+-n_- =\frac{1}{32\pi^2}\int_X d^4x \varepsilon^{\mu\nu\rho\sigma}
{\rm tr}F_{\mu\nu}F_{\rho\sigma}(x),
\end{align}
where $n_\pm$ is the number of zero modes with $\pm$ chirality.

In this section, we have derived the axial $U(1)$ anomaly
and Atiyah-Singer index theorem with the Pauli-Villars regularization.
We have confirmed that the axial $U(1)$ anomaly is an explicit breaking
of the theory, which comes from the mass term of the PV field.
The RHS of the index theorem is the winding number of the gauge fields,
while the LHS is the index of the Dirac operator.
The LHS is given by the zero modes only, 
which is a ``bonus'' from a property of the Dirac operator
$\{\gamma_5, D\}=0$. In the following, we
will discuss what we can do when this bonus is missing.

\section{Atiyah-Singer index with massive Dirac operator}
\label{sec:AS}

Let us consider a massive fermion action from the beginning,
\begin{align}
  S[m,\theta] = \int_X d^4x\; \bar{\psi}\left[D+m\exp(i\gamma_5 \theta)\right]\psi(x) + \int_X d^4x\; \bar{\phi}(D+M)\phi(x),
\end{align}
where we assign a nontrivial chiral angle $\theta$ to the mass term.
By a chiral rotation, the path integral is equal to
\begin{align}
  \label{eq:Zm}
Z[m,\theta]&=\int [D\bar{\psi}D\psi]\int [D\bar{\phi}D\phi]e^{-S[m,\theta]}\nonumber \\
&= \int [D\bar{\psi'}D\psi']\textcolor{black}{J}\int [D\bar{\phi}'D\phi']\textcolor{black}{J^{-1}}
e^{-S[\textcolor{black}{m,0}]+\int_X d^4x \left[\bar{\phi}'\textcolor{black}{M (e^{i\gamma_5 \theta}-1)}\phi'(x)\right]}\nonumber\\
&=  \int [D\bar{\psi'}D\psi']\int [D\bar{\phi}'D\phi']
e^{-S[\textcolor{black}{m},0]+i\theta \int_X d^4x \left[\textcolor{black}{\frac{1}{32\pi^2}\varepsilon^{\mu\nu\rho\sigma}
      {\rm tr}F_{\mu\nu}F_{\rho\sigma}(x)}\right]}\nonumber\\
&=\frac{\det(D+m)}{\det(D+M)}e^{i\theta \int_X d^4x \left[\textcolor{black}{\frac{1}{32\pi^2}\varepsilon^{\mu\nu\rho\sigma}
{\rm tr}F_{\mu\nu}F_{\rho\sigma}(x)}\right]}.
\end{align}
For $m=0$, $\theta$ is unphysical being rotated away from the theory\footnote{
  Setting the up quark mass exactly zero was considered as a solution to the strong CP problem in QCD but
  the possibility was excluded by recent lattice QCD results\cite{FlavourLatticeAveragingGroup:2019iem}:
  up quark mass is different from zero by 25 standard deviations.
}.
For $m=M$, the fermion is completely decoupled from the theory and its
effect is renormalized into the $\theta$ term.

For a general value of $\theta$, the time-reversal $T$ or parity symmetry is broken.
At $\theta=0$, the system is $T$ symmetric but trivial,
which is expected in a normal insulator.
At $\theta=0$, the $T$ invariance is maintained in a nontrivial way,
\begin{align}
Z[m,\theta=\pi]&\propto(-1)^{I_{AS}}=(-1)^{-I_{AS}},
\end{align}
where $I_{AS}$ is the winding number or the RHS of the Atiyah-Singer index.
We will see below that $\theta=\pi$ represents physics of the topological insulator.

Note in this section that we have not introduced any notion of chiral or zero modes.
Nevertheless, the index is hidden in the massive fermion determinant.
Our proposal is then to use the massive Dirac fermion to ``define'' the index,
using anomaly rather than symmetry.

Let us rewrite the determinant above setting $m=M$ and $\theta=\pi$ as below.
\begin{align}
  Z[M,\pi]&=\frac{\det(D-M)}{\det(D+M)}=\frac{\det i\gamma_5(D-M)}{\det i\gamma_5(D+M)}
  = \frac{\prod_{\lambda_{-M}} i\lambda_{-M}}{\prod_{\lambda_{+M}} i\lambda_{+M}}\nonumber\\
  &= \exp\left[-\frac{i\pi}{2}\left(\sum_{\lambda_{+M}} {\rm sgn} \lambda_{+M}
    -\sum_{\lambda_{-M}} {\rm sgn} \lambda_{-M} \right)\right],
\end{align}
where $\lambda_\pm$ are the eigenvalues of $H_\pm = \gamma_5(D\pm M)$.
The exponent contains the so-called APS $\eta$ invariant
defined by the summation of sign of the eigenvalues,
and the ``new'' definition
of the Atiyah-Singer index is given as
\begin{align}
  \label{eq:ASeta}
  I_{AS} &= - \frac{1}{2}\eta(H_-) + \frac{1}{2}\eta(H_+) =: - \frac{1}{2}\eta(H_-)^{\rm PV.},
\end{align}
where the last equality reminds us that the second term is contribution
from the PV fields\footnote{
  In the original definition in Ref.~\citen{Atiyah:1976jg}
  the $\eta$ invariant was given by the $\zeta$ function regularization.
  Here we use the Pauli-Villars but we may consider each
  term of Eq.~(\ref{eq:ASeta}) is regularized by the $\zeta$ function before
  taking the difference.
  In either case, the result is the same.
}.
This definition does not need chiral symmetry or $\{D,\gamma_5\}=0$.
Instead, it is no longer given by the zero modes only.

\subsection{Perturbative computation}

In order to directly confirm that the RHS of Eq.~(\ref{eq:ASeta}) is equivalent
to the Atiyah-Singer index, let us express the $\eta$ invariant
in an integral form of a ``half'' Gaussian:
\begin{align}
\eta(H)={\rm Tr}\frac{H}{\sqrt{H^2}}
= \frac{2}{\sqrt{\pi}}\int_0^\infty du\;{\rm Tr} H e^{-u^2 H^2},
\end{align}
and perform a weak coupling expansion in $e^{-u^2 H^2}$.
Note for very large eigenvalue of $H$ that
the short distance behavior $u\sim 0$ is subtle to evaluate.
But such a possible UV divergence is precisely canceled
between $H_\pm$ contributions in Eq.~(\ref{eq:ASeta}).

Using $H_\pm^2 = \gamma_5(D\pm M)\gamma_5(D\pm M)= -D^2+M^2$, we have
\begin{align}
  \eta(H_\pm) &= \pm \frac{2M}{\sqrt{\pi}}\int_0^\infty du\;{\rm Tr}\textcolor{black}{\gamma_5}(1\pm D/M) e^{-u^2 (-\textcolor{black}{D^2}+M^2)}\nonumber\\
   &= \pm \frac{2M}{\sqrt{\pi}}\int_0^\infty du\;e^{-u^2M^2}{\rm Tr}\textcolor{black}{\gamma_5}e^{u^2 \textcolor{black}{D^2}}
  \nonumber\\
  &=\frac{\pm 1}{32\pi^2}\int d^4x\;\;\varepsilon^{\mu\nu\rho\sigma}
{\rm tr}F_{\mu\nu}F_{\rho\sigma}(x)+\mathcal{O}(1/M),
\end{align}
where we have used that $D^2$ contains only even number of $\gamma_\mu$ products.
The evaluation of ${\rm Tr}\textcolor{black}{\gamma_5}e^{u^2 \textcolor{black}{D^2}}$
is exactly the same as the standard Fujikawa method with the heat kernel regularization.
Now we have perturbatively confirmed
\begin{align}
  \label{eq:ASeta2}
  - \frac{1}{2}\eta(H_-)^{\rm PV.} = \frac{1}{32\pi^2}\int d^4x\;\;\varepsilon^{\mu\nu\rho\sigma}
{\rm tr}F_{\mu\nu}F_{\rho\sigma}(x),
\end{align}
at least, in the large $M$ limit.
The reader may wonder why this $\eta$ invariant is an integer,
whereas for a general Hermitian operator $h$, $\eta(h)$ is known to be a non-integer.
It is the time-reversal symmetry of the fermion determinant
in four dimensions that
guarantees that $\eta(H_-)^{\rm PV.}/2$ is always an integer.

\subsection{A nonperturbative proof}
\label{subsec:ASproof}

If we use the chiral symmetry of $D$,
we can easily give a nonperturbative proof for the equality (\ref{eq:ASeta})
with a finite $M$.
From the anti-commutation relation $\{H_\pm,D\}=0$,
it is guaranteed that every eigenmodes of $H_\pm$ appear in $\pm$ pairs,
except for the zero modes of $D$ (or $\pm M$ modes of $H_\pm$). 
Then we have
\begin{align}
\eta(H_\pm) = \pm (n_+-n_-) + (\mbox{cutoff effects}),
\end{align}
where a possible regularization dependent part $(\mbox{cutoff effects})$
is precisely canceled by the difference in Eq.~(\ref{eq:ASeta}).

We, however, would like to try another proof (we showed  in our paper \cite{Fukaya:2019qlf}),
which is a bit tedious but useful in the application to
the APS index where we do not have any nice anti-commutation relation.
The new proof is also physically interesting
as it introduces a similar structure to topological insulator.

\begin{figure}[b]
\centerline{\includegraphics[width=12cm]{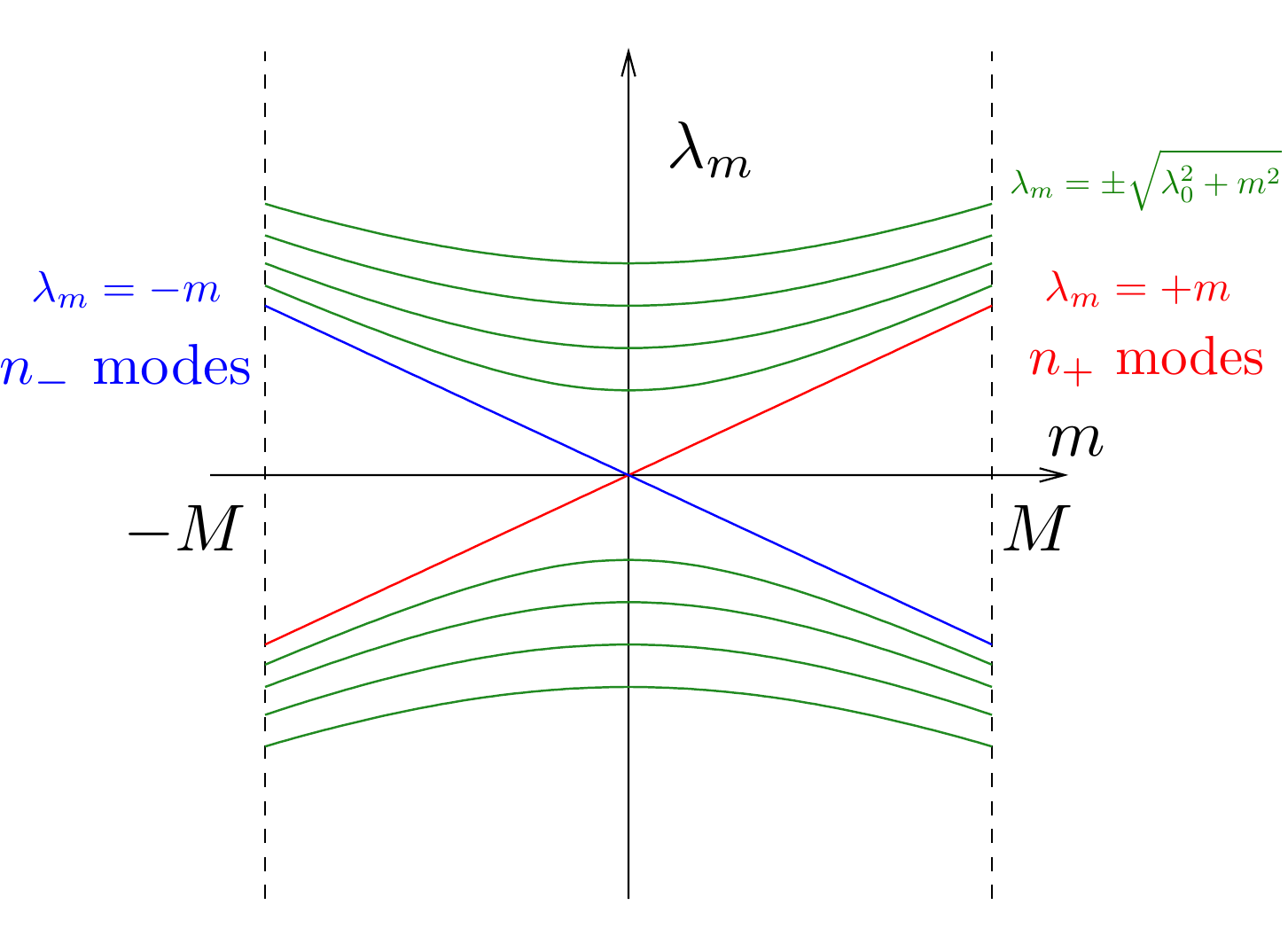}}
\caption{
The spectrum of $H(m)=\gamma_5(D+m)$ as a function of $m$.
  \label{fig:ASspecflow}}
\end{figure}

To this end, let us treat $H(m)=\gamma_5(D+m)$ as a one parameter family
in the range $-M \leq m \leq M$.
For the zero modes $D\phi=0$, $H(m)\phi=\pm m\phi$, where
the sign is equal to  the chirality.
For the nonzero modes of $D$, every eigenmode of $H(m)$ appear in $\pm$ pairs
with $\lambda_m = \pm \sqrt{\lambda_0^2+m^2}$, where $\lambda_0$ is
one of the pair at $m=0$.
The spectrum of $\{\lambda_m\}$ is shown in Fig.~\ref{fig:ASspecflow}.
It is important to note that $n_+$ eigenmodes cross zero from
negative to positive, and $n_-$ cross from positive to negative, respectively.
In mathematics, the difference of crossings
$n_+-n_-$ is known as the spectral flow of $H(m),\;m\in [-M,+M]$.

In Fig.~\ref{fig:ASspecflow}, we may consider the mass parameter as
an extra coordinate $x_5=m$. Then we notice two domains $x_5=m<0$
and $x_5=m>0$ where the fermion has a gap.
At $x_5=m=0$, the gap closes and the fermion system has
chiral symmetry. This structure reminds us of the domain-wall fermion,
or topological insulator.
The result is not special for
the linear function $m=x_5$ but unchanged for any monotonically increasing
function $m(x_5)$ keeping $m(0)=0$. 

In fact, our new proof\cite{Fukaya:2019qlf} for Eq.~(\ref{eq:ASeta})
starts with constructing
the domain-wall fermion, taking a kink mass $m(x_5)=M{\rm sgn}(x_5)$,
where $x_5 \in \mathbb{R}$ is the fifth coordinate
on a cylindrical manifold $X \times \mathbb{R}$.
On this five-dimensional manifold, we introduce a two-flavor
Dirac fermion and a Dirac operator
\begin{align}
D_{\rm DW}^{\rm 5D}(M)&=\tau_2 \otimes\partial_{x_5} + i\tau_1 \otimes \gamma_5(D+M{\rm sgn}(x_5))\nonumber\\
&=i\left(
\begin{array}{cc}
0 & \partial_{x_5}+\gamma_5(D+M{\rm sgn}(x_5))\\
-\partial_{x_5}+\gamma_5(D+M{\rm sgn}(x_5)) & 0
\end{array}
\right),
\end{align}
setting $A_5=0$ and $\partial_{x_5}A_{\mu=1,2,3,4}=0$.
This Dirac operator can be also viewed as a massless
six-dimensional single-flavor operator with
the 6th momentum fixed to $M{\rm sgn}(x_5)$.
In either interpretation, the Dirac operator has a ``chiral'' symmetry
with $\gamma_7 = \tau_3 \otimes 1_{4\times 4}$.

Now let us solve the Dirac equation,
\begin{align}
  \label{eq:DWeq}
D_{\rm DW}^{\rm 5D}(M)\phi(x)&=\left[\tau_2 \otimes\partial_{x_5} + i\tau_1 \otimes \gamma_5(D+M{\rm sgn}(x_5))\right]\phi(x)=0.
\end{align}
In the large $M$ limit, any zero mode must satisfy
\begin{align}
  \label{eq:DWsol}
\tau_2 [1\otimes\partial_{x_5} + \tau_3 \otimes \gamma_5 M{\rm sgn}(x_5)]\phi(x)=0,
\end{align}
to which an edge-localized solution
\begin{align}
\phi(x)\propto \exp(-M|x_5|),\;\;\;\tau_3\otimes\gamma_5 \phi(x)=+ \phi(x)
\end{align}
is known  \cite{Jackiw:1975fn, Callan:1984sa},
and the massless Dirac equation $D\phi(x)=0$.
Since $\gamma_7 = \tau_3 \otimes 1_{4\times 4}$ and $1_{2\times 2}\otimes \gamma_5$
must have the same eigenvalue, we can conclude that
their indices are equal:
\begin{align}
  {\rm Ind}(D_{\rm DW}^{\rm 5D}(M)) ={\rm Ind}(D)\;\;
  \left( \lim_{s\to0}{\rm Tr}\gamma_7e^{sD_{\rm DW}^{\rm 5D}(M)^2}=
  \lim_{s\to0} {\rm Tr}\gamma_5e^{sD^2}\right).
\end{align}
In mathematics, this is known as the localization (and product formula)
of the zero modes\cite{Witten:1982im, FurutaIndex}. With the position dependent mass,
we can make the zero modes localized on a lower-dimensional surface
and the index is given by the product of the one
in the lower dimensions, and another in the normal directions.
In our case, the index in the normal direction corresponds
to the solution to Eq.~(\ref{eq:DWsol}), which is always unity.

Let us take a different view of the same Dirac equation (\ref{eq:DWeq}).
At $x_5\sim -\infty$, the equation is
\begin{align}
0
&= \tau_2 \left[1 \otimes\partial_{x_5} + \tau_3 \otimes H(-M)\right]\phi(x).
\end{align}
For the chirality $\gamma_7 = \tau_3 \otimes 1_{4\times 4}=\pm 1$,
we find solutions of the form
\begin{align}
\phi \sim \exp(\mp \lambda_{-M}x_5),
\end{align}
where $\lambda_{-M}$ is the eigenvalue of $H(-M)$,
which is normalizable only when ${\rm sgn}(\lambda_{-M}) = \mp 1$.

Similarly, at $x_5\sim +\infty$, the equation 
\begin{align}
0
&= \tau_2 \left[1 \otimes\partial_{x_5} + \tau_3 \otimes H(+M)\right]\phi(x)
\end{align}
has the solutions for $\gamma_7 = \tau_3 \otimes 1_{4\times 4}=\pm 1$
with the eigenvalue $\lambda_{+M}$ of $H(+M)$ such that
\begin{align}
\phi \sim \exp(\mp \lambda_{+M}x_5),
\end{align}
which is normalizable only when ${\rm sgn}(\lambda_{+M}) = \pm 1$.

Smoothing the step function\footnote{
  In our paper \cite{Fukaya:2019qlf},
  we did not use this intuitive adiabatic approach
  but gave a different proposition valid
  for arbitrarily steep $x_5$ dependence of $m(x_5)$.  
}
and considering an adiabatic
$x_5$ dependence of the solutions, we can
assign a one-to-one correspondence between the solution
and the one-parameter family of each eigenvalue $\lambda_m$.
For a $\gamma_7=+1$ zero mode,
we have to find a $\lambda_m$ with $\lambda_{-M}<0$
and $\lambda_{+M}>0$, while for $\gamma_7=-1$
we have to find the one with $\lambda_{-M}>0$ and $\lambda_{+M}<0$.
Namely, the index is given by the spectral flow along the path $m\in [-M,+M]$
counting the increased number of the positive eigenvalues
subtracted by that of the negative eigenvalues divided by 2:
\begin{align}
  {\rm Ind}(D_{\rm DW}^{\rm 5D}(M))&= \frac{1}{2}\left(\sum_{\lambda_{+M}>0}-\sum_{\lambda_{-M}>0}-\sum_{\lambda_{+M}<0}+\sum_{\lambda_{-M}<0}\right)
  \nonumber\\
&=\frac{1}{2}\sum_{\lambda_{+M}}{\rm sgn}(\lambda_{+M})-\frac{1}{2}\sum_{\lambda_{-M}}{\rm sgn}(\lambda_{-M}).
\end{align}
As the last equality leads to the difference of the eta invariants\footnote{
  We have implicitly used a fact that the index is given by the eta invariants
  at $x_5=\pm \infty$, which is true in odd dimensions only.
  },
the proof for Eq.(\ref{eq:ASeta}) is complete.

In this proof, the original four-dimensional manifold
appears as if it were a domain-wall between five-dimensional
topological insulator in the $x_5<0$ region and
normal insulator in the $x_5>0$ region.
While the standard index of massless Dirac operator
is given on the edge or domain-wall,
the massive expression corresponds to
the evaluation of the same quantity at the bulk.
We may call it a ``bulk-edge correspondence''.

\begin{figure}[b]
\centerline{\includegraphics[width=10cm]{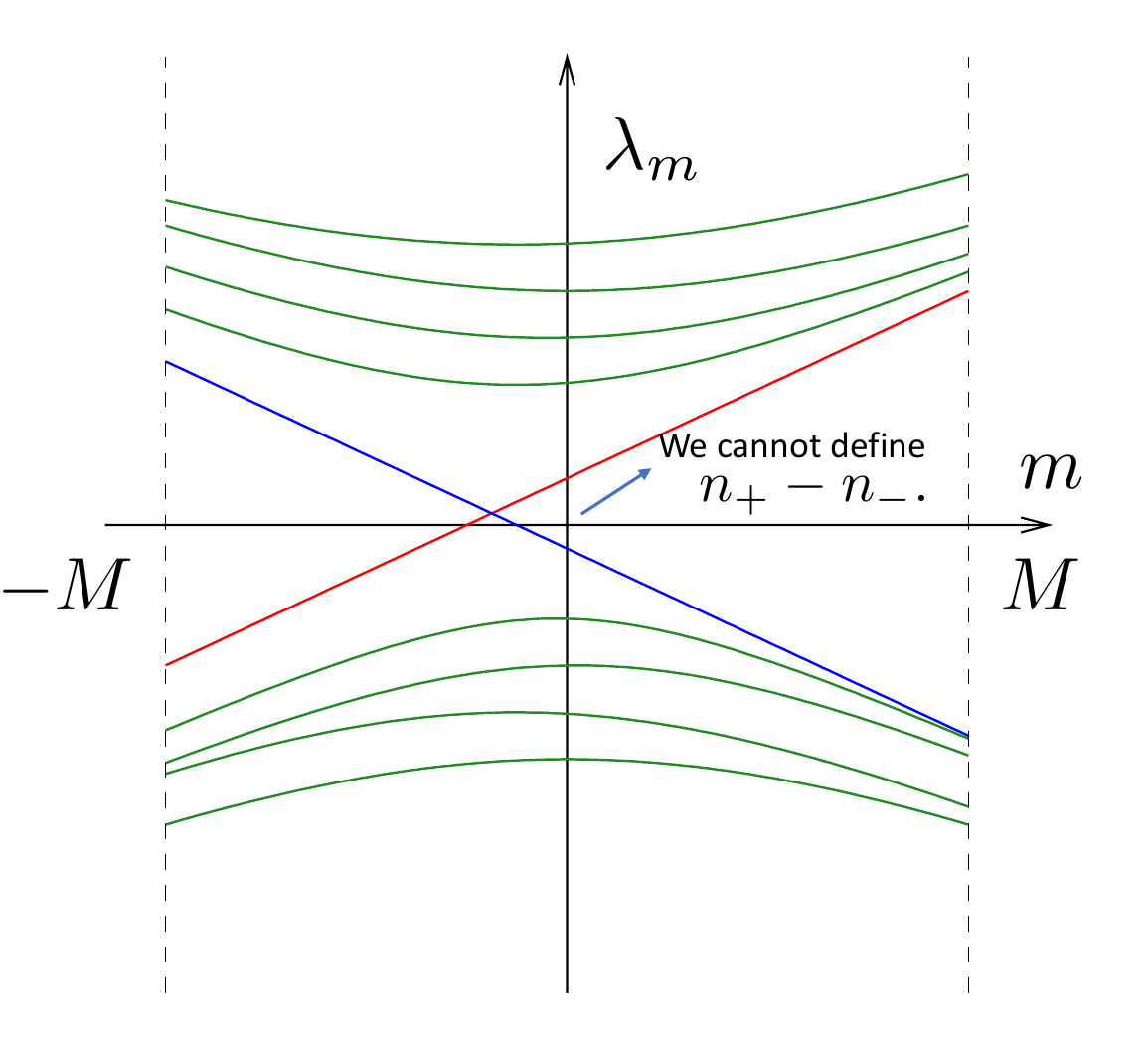}}
\caption{
  Example of the spectrum of $H(m)=\gamma_5(D+m)$
  when $D$ does not respect the chiral symmetry.
  The massive definition or the number of crossings is still easy to count,
  while it is difficult to find a good definition of the chiral zero modes at $m=0$.
  \label{fig:ASspecflow2}}
\end{figure}

Before concluding this section,
we would like to give two remarks.
The first one is about stability of the definitions.
Suppose that we choose a regularization of the theory
where the chiral symmetry of the massless Dirac operator is explicitly lost,
which regularly happens on a lattice.
Then the spectrum may be distorted like Fig.~\ref{fig:ASspecflow2}.
It would be difficult to define the chiral zero modes,
  while it is not difficult to count
  the spectral flow as far as $H(\pm M)$ are gapped.
The massive definition is more stable against the symmetry breaking. 
  The second remark is a relation to K-theory.
  The equality of the two definitions of the index
  is a mathematical consequence of the so-called suspension isomorphism
  between a K group with the $Z_2$ grading chirality operator
  and another without chirality operator on a higher-dimensional
  manifold called the reduced suspension of the original one.

To summarize this section, we have argued that the Atiyah-Singer index
can be described by the eta invariant of
the massive Dirac fermion operator without using the chiral symmetry.
The equality has been confirmed both perturbatively and nonperturbatively.
The nonperturbative proof utilizes a structure similar to
a five-dimensional topological insulator and
the equality is established by a bulk-edge correspondence.
The massive bulk definition is stable against symmetry breaking
of the original Dirac operator.

\section{Atiyah-Patodi-Singer(APS) index (review)}
\label{sec:APSreview}

So far, we have discussed fermions on a closed manifold
without boundary.
In this section, we consider a manifold with boundary
and review the original Atiyah-Patodi-Singer index theorem
and discuss why the formulation is physicist-unfriendly.

For simplicity, we consider a four-dimensional flat Euclidean compact
manifold $X$ with a three-dimensional boundary $Y$.
Let $D_{\rm APS}$ be a Dirac operator for
the fermion field on $X$, to which the so-called
APS boundary condition is imposed.
Then the APS index theorem is
\begin{align}
  \label{eq:APS4D}
{\rm Ind}D_{\rm APS} = \frac{1}{32\pi^2}\int_X d^4x\;\;\varepsilon^{\mu\nu\rho\sigma}
{\rm tr}F_{\mu\nu}F_{\rho\sigma}(x)-\frac{1}{2}\eta(iD_Y),
\end{align}
where $iD_Y$ is a boundary Dirac operator on $Y$.
As $Y$ is an odd-dimensional manifold, the eta invariant
$\eta(iD_Y)$ is a non-integer in general,
which compensates the surface contribution of the first term
to make the total RHS an integer.

This APS index theorem has not been that relevant
in physics, since we were not interested in manifolds with boundary,
which breaks the Lorentz invariance.
However, recently, it attracts attention as pointed out in Ref.\citen{Witten:2015aba}
that the APS index is a key to understand the bulk-edge correspondence
of topological insulators.
In a topological insulator, the electrons in bulk is gapped, while
the edge or surface modes become massless showing a good conductivity. 
The first term of Eq.~(\ref{eq:APS4D}) corresponds to
the phase of bulk fermions, while the second term
is that of edge modes.
Each of them has anomaly in time-reversal $T$ symmetry,
but the total contribution makes the theory  $T$ invariant.

Let us look into the details. The massless edge mode's
determinant is
\begin{align}
Z_{\rm edge} &= \lim_{\mu\to \infty} \det\frac{D_Y}{D_Y+\mu}
=\lim_{\mu\to \infty} \prod_{\lambda_Y}\frac{i\lambda_Y}{i\lambda_Y+\mu}
\sim\lim_{\mu\to \infty} \prod_{\lambda_Y}\frac{i\lambda_Y}{\mu}\nonumber\\
&\propto \exp\left[-\frac{i\pi}{2} \sum_{\lambda_Y} {\rm sgn}(-\lambda_Y)\right]
= \exp\left[-\frac{i\pi}{2} \eta(iD_Y)\right].
\end{align}
For the bulk massive fermions, from the results on a closed manifold
we discussed in Eq.~(\ref{eq:Zm}) setting $m=M$ and $\theta=\pi$
it is natural to assume that
\begin{align}
Z_{\rm bulk}
&\propto \exp\left(i\pi \int_X d^4x \left[\textcolor{black}{\frac{1}{32\pi^2}\varepsilon^{\mu\nu\rho\sigma}
{\rm tr}F_{\mu\nu}F_{\rho\sigma}(x)}\right]\right),
\end{align}
holds even when $X$ has a boundary. The total partition function is then
\begin{align}
Z_{\rm bulk}Z_{\rm edge}\propto \exp(i\pi {\rm Ind}D_{\rm APS}) = (-1)^{{\rm Ind}D_{\rm APS}},
\end{align}
which is $T$ invariant under the change of ${\rm Ind}D_{\rm APS}\to -{\rm Ind}D_{\rm APS}$.

In this way, the RHS of Eq.~(\ref{eq:APS4D}) has a natural interpretation in physics.
But if you look at the definition of LHS, we would feel a bit uncomfortable.

Let us take our coordinate $x_4$ in the normal direction of the boundary $Y$
which is located at $x_4=0$.
The Dirac operator near $x_4=0$ is written as
\begin{align}
  D_{\rm APS} = \gamma_4\left(\frac{\partial}{\partial x_4}+A\right),
\end{align}
where $A=\gamma_4\gamma_iD^i$ is the three-dimensional operator
at $x_4=0$. The APS boundary condition requires the fermion fields
to kill all the positive eigenmode components of $A$:
\begin{align}
\label{eq:APSbc}
(A+|A|)\psi|_{x_4=0}=0,\;\;\;(A+|A|)D_{\rm APS}\psi|_{x_4=0}=0.
\end{align}
(For simplicity, we assume $A$ has no zero eigenvalue).
This condition is nonlocal, as clear from a nonlocal operator $|A|$,
and sensitive to the eigenfunctions of $A$, which is extended in $Y$. 

The APS boundary condition keeps the anti-Hermiticity of
$D_{\rm APS}$
\begin{align}
0 &= \int_{x_4>0} d^4x \phi_2^\dagger(x) D_{\rm APS}\phi_1(x) + \int_{x_4>0} d^4x (D_{\rm APS}\phi_2)^\dagger(x)\phi_1(x)
\nonumber\\
&= \int_{x_4=0} d^3x \phi_2^\dagger(x) \gamma_4 \phi_1(x),
\end{align}
since $\{\gamma_4, A\}=0$, and for any linear combination $\phi_1(x)$
of negative eigenmodes of $A$, $\gamma_4\phi_1(x)$ is that of positive eigenmodes.
The condition also keeps the chirality $[\gamma_5, A]=0$ well-defined,
and with the anti-commutation relation $\{\gamma_5, D_{\rm APS}\}$,
the index can be written by the chiral zero modes, as usual: $n_+-n_-$.

In the standard Fujikawa method, let us perturbatively evaluate the index,
with a heat kernel regulator,
\begin{align}
{\rm Ind}D_{\rm APS} = \lim_{s\to 0}{\rm Tr}\gamma_5e^{sD_{\rm APS}^2}.
\end{align}
Unlike on a closed manifold, the trace must be evaluated with a
complete set satisfying the APS boundary condition.
In the chiral representation,
\begin{align}
A=\gamma_4\gamma_iD^i=\left(
\begin{array}{cc}
-i\sigma_i D^i &0\\
0 & i\sigma_iD^i 
\end{array}
\right)=:
\left(
\begin{array}{cc}
iD_Y &0\\
0 & -iD_Y
\end{array}
\right) = \tau_3\otimes iD_Y
\end{align}
and we can take a basis of the form below.
\begin{align}
\phi_{\pm}^{\omega,\lambda}(x_4)\otimes\phi_\lambda^{Y}(\bm{x}),
\end{align}
where the subscript $\pm$ denotes the chirality:
$\tau_3\phi_{\pm}^{\omega,\lambda}(x_4)=\pm \phi_{\pm}^{\omega,\lambda}(x_4)$,
$\omega$ is the (absolute value of) momentum in the $x_4$ direction,
and $\phi_\lambda^{Y}(\bm{x})$ is the eigenfunction of the surface
Dirac operator $iD_Y$ (at $x_4=0$) with the eigenvalue $\lambda$:
$iD_Y\phi_\lambda^{Y}(\bm{x})=\lambda\phi_\lambda^{Y}(\bm{x})$.

The APS boundary condition reads
\begin{align}
  \phi_{+}^{\omega,\lambda}(x_4=0)&=0,\;\;\;
  \left.(\partial_4-\lambda)\phi_{-}^{\omega,\lambda}(x_4)\right|_{x_4=0}=0,\;\;\; \mbox{for }\lambda>0,\nonumber\\
  \phi_{-}^{\omega,\lambda}(x_4=0)&=0,\;\;\;
  \left.(\partial_4+\lambda)\phi_{+}^{\omega,\lambda}(x_4)\right|_{x_4=0}=0,\;\;\; \mbox{for }\lambda<0,
\end{align}  
to which the solutions are given by
\begin{align}
  \phi_{+}^{\omega,\lambda}(x_4)&=\frac{e^{i\omega x_4}-e^{-i\omega x_4}}{\sqrt{2\pi}},\;
  \phi_{-}^{\omega,\lambda}(x_4)=
  \frac{(i\omega+\lambda)e^{i\omega x_4}+(i\omega-\lambda)e^{-i\omega x_4}}{\sqrt{2\pi(\omega^2+\lambda^2)}}
  \;\;\;\mbox{for }\lambda>0,\nonumber\\
  \phi_{-}^{\omega,\lambda}(x_4)&=\frac{e^{i\omega x_4}-e^{-i\omega x_4}}{\sqrt{2\pi}},\;
  \phi_{+}^{\omega,\lambda}(x_4)=
  \frac{(i\omega-\lambda)e^{i\omega x_4}+(i\omega+\lambda)e^{-i\omega x_4}}{\sqrt{2\pi(\omega^2+\lambda^2)}}
  \;\;\;\mbox{for }\lambda<0,
\end{align}
where the eigenvalue of $D_{\rm APS}$ is $\pm i\sqrt{\lambda^2+\omega^2}$.

Interestingly, there is no edge-localized modes in the complete set.
In fact, the Dirac equation $D_{\rm APS}\phi=\gamma_4(\partial_4+A)\phi=0$
has a formal solution $\exp(-\lambda x_4)\otimes\phi_\lambda^{Y}(\bm{x})$
for any eigenvalue $\lambda$ of $A$, but the APS condition
does not allow normalizable solutions with $\lambda>0$, which decays at $x_4=\infty$.

With the above complete set, we are now ready to compute the index perturbatively.
At the leading order of the adiabatic expansion, ignoring the $x_4$ dependence
of the gauge field, we have
\begin{align}
  \lim_{s\to 0}{\rm Tr}\gamma_5e^{sD_{\rm APS}^2}|_{LO}
  &= \lim_{s\to 0}\sum_\lambda \int dx_4 {\rm sgn}\lambda e^{-s\lambda^2}\int\frac{d\omega}{2\pi}
  \left(-1+\frac{2i|\lambda|}{\omega+i|\lambda|}e^{-s\omega^2+2i\omega x_4}\right)\nonumber\\
  &= \lim_{s\to 0} \sum_\lambda {\rm sgn}\lambda  \int_0^\infty dx_4 \frac{\partial}{\partial x_4}
  \left[\frac{1}{2}e^{2|\lambda|x_4}{\rm erfc}(x_4\sqrt{s}+|\lambda|\sqrt{s})\right]\nonumber\\
  &= -\frac{1}{2}\eta(iD_Y).
\end{align}
Here ${\rm erfc}(z)$ is the complementary error function,
\[
{\rm erfc}(x)=\frac{2}{\sqrt{\pi}}\int_z^\infty d\xi e^{-\xi^2},
\]
which satisfies ${\rm erfc}(0)=1$ and ${\rm erfc}(\infty)=0$.
From the next-to-leading order (NLO) contribution at bulk, we reproduce the curvature term,
\begin{align}
  \lim_{s\to 0}{\rm Tr}\gamma_5e^{sD_{\rm APS}^2}|_{NLO}
  &= \frac{1}{32\pi^2}\int_X d^4x\;\;\varepsilon^{\mu\nu\rho\sigma}
{\rm tr}F_{\mu\nu}F_{\rho\sigma}(x).
\end{align}
See Ref.~\citen{Alvarez-Gaume:1984zst} for the details.
Thus we perturbatively confirmed the APS index theorem\footnote{
  With more general setups, a simple derivation of the APS index theorem
  was recently given in Ref.~\cite{Kobayashi:2021jbn}.
}.

The APS condition has no problem in mathematics.
To the first-order differential equations, 
we can put any boundary condition by hand.
In physics, however, it is unnatural to keep helicity
of one particle state as described in Sec.~\ref{sec:intro}.
Also, in quantum field theory, the quantum correction
makes the boundary condition distorted, which could end up
with a natural boundary condition in the continuum limit \cite{Luscher:2006df}.
Specifically, the APS condition would receive a power
divergent correction (expected from a naive dimensional analysis),
\begin{align}
(A+|A|+c/a)\phi =0,
\end{align}
where $a$ is the short-range cut-off of the theory,
which would end up with a Dirichlet condition
in the continuum limit, unless we fine-tune the coefficient $c$,
or require it to vanish imposing some symmetry.
It is also clear that even if we could fine-tune $c$ to vanish,
the system with the APS boundary condition, under which
no edge localized mode can exist,
is far different from topological insulators.

As a final discussion of this section, let us consider
what is more physical setup.
In physics, anything we regard as a boundary has
``outside'' of it. Topological insulators are nontrivial
because their outside is covered by normal insulators (including air).
Therefore, it would be better to consider physics
with a domain-wall, separating two or more
regions with different physical properties, rather
than that on a manifold with a boundary.
As discussed in this section,
we should require angular momentum of particles to be preserved,
rather than their helicity.
The boundary condition should not be put by hand,
but should be automatically chosen by nature.
Hopefully the edge localized modes can exist
and play a key role.
In the next section, we will see that the domain-wall fermion
perfectly matches these requirements\footnote{
  In Ref.~\citen{Witten:2019bou}, why the APS condition
  appears in physics and its non-local behavior has no problem is differently explained.
  They rotate the $x_4$ to the “time” direction and introduce
  the APS boundary condition as an intermediate “state”.
  They showed that the unphysical property of APS is canceled between the bra/ket states.
  It is interesting to note that we try to remove it in our work, while
  in Ref.~\citen{Witten:2019bou} they try to cancel it.
}.

\section{APS index with domain-wall fermion}
\label{sec:APSDW}

In this section, we show that the domain-wall Dirac operator
can describe the APS index with different setup from the original one.

Let us recall Sec.~\ref{sec:AS},
where we have confirmed that the AS index $I_{\rm AS}$
can be expressed by the massive Dirac operator, which is reflected in
its determinant,
\begin{align}
\det\left(\frac{D-M}{D+M}\right) \propto (-1)^{\textcolor{black}{I_{\rm AS}}}.
\end{align}
In this section, we consider a four-dimensional domain-wall fermion determinant,
\begin{align}
\det\left(\frac{D-\varepsilon M}{D+M}\right),
\end{align}
where $\varepsilon$ denotes a position-dependent step function, for example, $\varepsilon = {\rm sgn}(x_4)$\footnote{
  The index and anomaly with more general position-dependent mass term
  was recently discussed in Ref.~\citen{Kanno:2021bze}.
}.

From the $\gamma_5$ Hermiticity: $\gamma_5 D \gamma_5 = -D = D^\dagger$,
we find that the  domain-wall fermion determinant is real,
and we may assign an integer $I$
\begin{align}
\det\left(\frac{D-\varepsilon M}{D+M}\right) 
&= \det\left(\frac{\gamma_5(D-\varepsilon M)\gamma_5}{\gamma_5(D+M)\gamma_5}\right)
=\det\left(\frac{D-\varepsilon M}{D+M}\right)^* \propto (-1)^I,
\end{align}
to represent its sign. In the same way as in Sec.~\ref{sec:AS},
we can express $I$ by the $\eta$ invariants as follows,
\begin{align}
I=\frac{1}{2}\eta(H_{\rm DW})^{\rm PV.}=-\frac{1}{2}\eta(H_{\rm DW})+ \frac{1}{2}\eta(H_{\rm PV}),
\end{align}
where $H_{\rm DW}=\gamma_5(D-\varepsilon M)$ and $H_{\rm PV}=\gamma_5(D+ M)$.
What is $I$ here? In fact, it is the APS index of $\varepsilon=+1$ region as will be shown below.

\subsection{Perturbative evaluation of $\eta$ invariant of domain-wall Dirac operator}
Let us perturbatively evaluate the $\eta$ invariant of the domain-wall Dirac operator,
\begin{align}
  \frac{1}{2}\eta(H_{DW})&=
  \frac{1}{2}{\rm Tr}\frac{H_{DW}}{\sqrt{H_{DW}^2}}
=
{\rm Tr} \frac{H_{DW}}{\sqrt{\pi}}\int_0^\infty du e^{-u^2H^2_{DW}}.
\end{align}
The short distance behavior around $u\sim 0$
for the large eigenvalues of $H_{DW}$,
is again precisely canceled with that of the Pauli-Villars.
In contrast to the AS index, 
the standard plane wave set does not work in evaluating the trace,
since the translational symmetry is broken by the position-dependent mass term.

As an explicit example, we consider a flat four-dimensional torus of size $L^3\times 2T$,
on which we put two domain-walls at $x_4=0$ and $x_4=T$.
Taking the $A_4=0$ gauge, the domain-wall Dirac operator is given by
\begin{align}
H_{\rm DW}=\gamma_5\left[\gamma_iD^i+ \gamma_4\partial_4- M{\rm sgn}(x_4){\rm sgn}(T-x_4)\right].
\end{align}

Near $x_4=0$ and taking the large $T$ limit, the solutions to
\begin{equation}
[H^{\rm free}_{\rm DW}]^2 \phi= \left[-\partial_\mu^2+M^2+2M\gamma_4\delta(x_4)\right] = \lambda^2\phi
\end{equation}
are $\varphi^{\omega/{\rm edge}}_{\pm,e/o}(x_4)\otimes e^{i\bm{p}\cdot\bm{x}}$, where the bulk modes 
\begin{align}
  \varphi^\omega_{\pm, o}(x_4)&=\frac{e^{i\omega x_4}-e^{-i\omega x_4}}{\sqrt{2\pi}},
  \varphi^{\omega}_{\pm,e}(x_4)=
  \frac{(i\omega\pm M)e^{i\omega |x_4|}+(i\omega\mp M)e^{-i\omega |x_4|}}{\sqrt{2\pi(\omega^2+M^2)}}
\end{align}
have eigenvalues $\lambda^2=\bm{p}^2+\omega^2+M^2$,
and 
\begin{align}
  \varphi^{\rm edge}_{-, e}(x_4)&=\sqrt{M}e^{-M|x_4|},
\end{align}
are the chiral edge localized modes with  $\lambda^2=\bm{p}^2$.
Note here that the subscript $\pm$ indicates the eigenvalue of $\gamma_4$,
and those with $e/o$ are even/odd under the reflection $x_4\to -x_4$.
The edge modes appear only in the $\gamma_4=-1$ and even sector.

It is important to note that we did not put any boundary condition by hand.
Nevertheless, they automatically satisfy 
\begin{align}
  \left[\partial_4 \mp M\varepsilon\right]\varphi^{\omega/{\rm edge}}_{\pm,e}(x_4)|_{x_4=0} = 0,\;\;\;
  \varphi^{\omega}_{\pm,o}(x_4=0)=0,
\end{align}
due to the domain-wall. More importantly, this condition is
local and preserves angular-momentum in the $x_4$ direction, but
does not keep the chirality.

For the edge modes, the domain-wall fermion Dirac operator
acts as
\begin{align}
  H_{DW}\phi^{edge}(x)&= \gamma_5(\gamma_iD^i +\underbrace{\gamma_4\partial_4-M\epsilon(x_4)}_{=0})e^{i\bm{p}\cdot \bm{x}}P^4_-\varphi_{-,e}^{edge}(x_4)\nonumber\\
  &=(\gamma_5\gamma_iP^4_-)D^i e^{i\bm{p}\cdot \bm{x}}\varphi_{-,e}^{edge}(x_4) \nonumber\\
  &=\left(\begin{array}{cc}
 0 & 0\\
0 & iD^{\rm 3D}
\end{array}\right)
 e^{i\bm{p}\cdot \bm{x}}\varphi_{-,e}^{edge}(x_4),
\end{align}
where $P^4_-=\frac{1-\gamma_4}{2}$, $iD^{\rm 3D}=-i\sigma_iD^i(x_4)$, and we have used notations
\begin{align}
  \gamma_i&=\tau_2\otimes\sigma_i,\;\gamma_4=\tau_3\otimes 1,\; \gamma_5=\tau_1\otimes 1,\nonumber\\
  \gamma_5\gamma_i P_-^4 &= (i\tau_3\otimes\sigma_i)P_-^4= -\frac{1-\tau_3}{2}\otimes i\sigma_i,
\end{align}
where $\sigma_i$ and $\tau_i$ are the Pauli matrices.

Then the edge mode's contribution to the $\eta$ invariant is
\begin{align}
-\frac{1}{2}\eta(H_{DW})^{edge}|_{x_4\sim 0} &= -\frac{1}{2}\sum_{edge modes}\phi^{edge}(x)^\dagger {\rm sgn}(H_{DW})\phi^{edge}(x)\nonumber\\
 &= -\frac{1}{2} \sum_{edge modes}\phi^{edge}(x)^\dagger \left[{\rm sgn}(iD^{\rm 3D}|_{x_4=0})+O(|x_4|)\right]\phi^{edge}(x)\nonumber\\
&= -\frac{1}{2}\eta(iD^{\rm 3D})|_{x_4=0}\times\left(\underbrace{\int dx_4 (\varphi^{edge}_{-,e})^\dagger \varphi^{edge}_{-,e}(x_4)}_{=1} + O(1/M)\right),
\end{align}
where we have used a fact that the $x_4$ dependence can be treated
as an expansion with respect to $1/M$ for the edge modes, 
\begin{align}
\int_{-\infty}^{+\infty} dx_4[ \varphi^{\rm edge}_{-,e}(x_4)]^\dagger x_4^n\varphi^{\rm edge}_{-,e}(x_4)&= M \int_{-\infty}^{+\infty} dx_4  x_4^n e^{-2M|x_4|}\nonumber\\
<  M \int_{-\infty}^{+\infty} dx_4  |x_4|^n e^{-2M|x_4|}&=2M \int_{0}^{+\infty} dx_4  x_4^n e^{-2M x_4} \nonumber\\
=2M(-1/2)^n\frac{\partial^n}{\partial M^n}\int_{0}^{+\infty} dx_4  e^{-2M x_4} 
&=2M(-1/2)^n\frac{\partial^n}{\partial M^n}\left(\frac{1}{2M}\right) \nonumber\\
&=\left\{
\begin{array}{cc}
1 & (n=0)\\
O(1/M^n) & (n>0)
\end{array}\right..
\end{align}

Adding contribution from another set of edge modes at $x_4=T$,
we have
\begin{align}
-\frac{1}{2}\eta(H_{DW})^{edge} 
&= -\frac{1}{2}\eta(iD^{\rm 3D})|_{x_4=0}+\frac{1}{2}\eta(iD^{\rm 3D})|_{x_4=T},
\end{align}
up to $1/M$ corrections. The sign of the second term is positive,
because the mass term changes the sign from negative to positive at $x_4=T$.

For the bulk modes, the eigenvalues satisfy $\lambda^2=\bm{p}^2+\omega^2+M^2>M^2$
and therefore, their contribution is local and it is easier to compute “density”,
\begin{align}
-\frac{1}{2}\eta(H_{DW})^{bulk}\textcolor{black}{(x)} 
&= -\frac{1}{2} \sum_{bulk modes}\phi^{bulk}(x)^\dagger \frac{H_{DW}}{\sqrt{H_{DW}^2}}\phi^{bulk}(x)\nonumber\\
&= -\sum_{bulk modes}\phi^{bulk}(x)^\dagger \frac{H_{DW}}{\sqrt{\pi}}\int_0^\infty du e^{-u^2H^2_{DW}}\phi^{bulk}(x)\nonumber\\
&=-\left.\frac{1}{\sqrt{\pi}}\sum_{bulk modes}\phi^{bulk}(x)^\dagger 
\right[\nonumber\\
 & \left.(\textcolor{black}{-\gamma_5 M\epsilon(x_4)}+\gamma_5 D)
\int_0^\infty du e^{-u^2(\lambda^2+\textcolor{black}{\gamma_{[\mu,\nu]}F_{\mu\nu}})}\right]
\phi^{bulk}(x)\nonumber\\
&= \frac{1}{64\pi^2} \textcolor{black}{\epsilon(x_4)}\epsilon_{\mu\nu\rho\sigma}{\rm tr}_cF^{\mu\nu}F^{\rho\sigma}(x) + O(1/M).
\end{align}
See Ref.~\citen{Fukaya:2017tsq} for the details.
An easier evaluation is to simply take the trace with
the plane wave complete set.
This should be valid as far as $x_4$ is larger than $1/M$,
where the boundary effect is exponentially small.
The Pauli-Villars contribution can be evaluated with the standard
plane waves,
\begin{align}
  \frac{1}{2}\eta(H_{PV})^{bulk}(x) &=\frac{1}{64\pi^2} \epsilon_{\mu\nu\rho\sigma}{\rm tr}_cF^{\mu\nu}F^{\rho\sigma}(x) + O(1/M).
\end{align}

Now the total index becomes
\begin{align}
  \label{eq:FOY}
  -\frac{1}{2}\eta(H_{\rm DW})+ \frac{1}{2}\eta(H_{\rm PV})
&= \frac{1}{32\pi^2}\int_{\textcolor{black}{0<x_4<T}} d^4x 
  \epsilon_{\mu\nu\rho\sigma}{\rm tr}_cF^{\mu\nu}F^{\rho\sigma}(x)
  \nonumber\\&-\frac{1}{2}\eta(iD^{\rm 3D})|_{x_4=0}+\frac{1}{2}\eta(iD^{\rm 3D})|_{x_4=T},
\end{align}
which coincides with the APS index on a cylinder $L^3\times [0,T]$.
Here we have neglected $O(1/M)$ corrections.
However, in Ref.~\citen{Fukaya:2017tsq}, we have shown that the left-hand side
of Eq.~(\ref{eq:FOY}) is an $M$-independent integer,
purely determined by the gauge field.

The boundary $\eta$ invariant represents the anomaly of
time-reversal symmetry of the edge-localized mode of topological insulator,
which is precisely canceled by that of the bulk modes.
The bulk and edge decomposition of the
domain-wall fermion Dirac operator matches this physical picture.

\subsection{A non-perturbative proof}

In the previous subsection, we have perturbatively
confirmed the equality
\begin{align}
  \label{eq:APSeta}
  {\rm Ind}D_{\rm APS} =-\frac{1}{2}\eta(H_{\rm DW})^{\rm PV.}
\end{align}
on a flat four-dimensional manifold.
The two quantities are defined on different manifolds.
The original APS index is given on a manifold with boundary,
while the $\eta$ invariant of the domain-wall Dirac operator
is given on a closed manifold.
The boundary conditions are also quite different.
The readers may wonder if the equality is true on
a general manifold or just a coincidence.
In Ref.~\citen{Fukaya:2019qlf} we derived
a theorem that for any APS index of a massless Dirac
operator on a even-dimensional
curved manifold $X_+$ with boundary, there exists a massive (domain-wall)
Dirac operator on a closed manifold, sharing its half with $X_+$,
and its $\eta$ invariant is equal to the original index.
In this subsection, we give a physicist-friendly sketch of
our mathematical proof.

\begin{figure}[bth]
\centerline{\includegraphics[width=12cm]{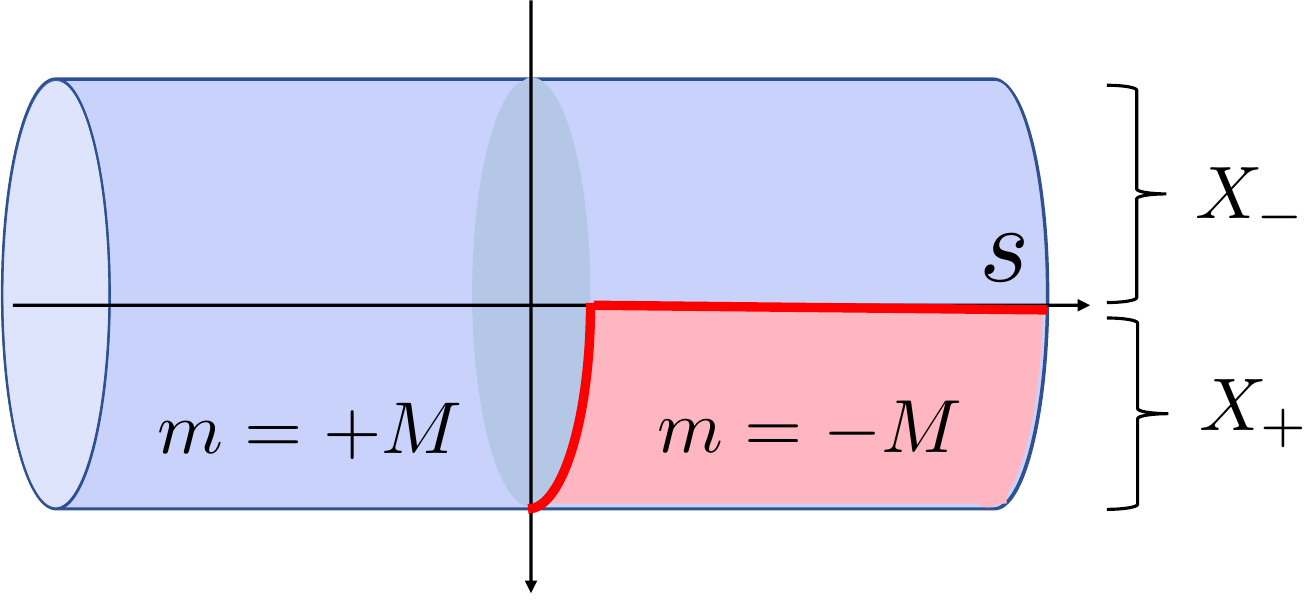}}
\caption{
  The cylindrical manifold $X\times \mathbb{R}$
  and the mass configuration.
  \label{fig:bentDW}}
\end{figure}

First, we introduce a direct product $X\times \mathbb{R}$,
where $X=X_+\cup X_-$ is a closed $2n$-dimensional manifold
on which the domain-wall fermion lives.
$X_\pm$ denotes the region where the mass term has $\pm$ sign. 
On this extended manifold, we modify the mass term as
\begin{align}
m(x,s)=\left\{
\begin{array}{cc}
-M & \mbox{for $s>0$, $x\in X_+$}\\
+M & \mbox{otherwise}
\end{array}
\right.,
\end{align}
as presented in Fig.~\ref{fig:bentDW}.
Here $x$ denotes a coordinate in $X$ and $s$ is in the $\mathbb{R}$ direction.
In this setup, we assume that the gauge field and metric on $X$ are just copied
in the $s$ direction, and the $s$ component of the gauge field is zero.
Note here that on the $s=+1$ slice, the mass term is the same as the original kink mass on $X$
and that on the $s=-1$ slice is positive everywhere in $X$ as in the Dirac operator on the
Pauli-Villars field.
Compared to the AS index case we discussed in Sec.~\ref{subsec:ASproof},
the domain-wall located at $s=0$ (or $x_5=0$) is bent and extended
towards the $s=+\infty$ direction.

In a similar way as in Sec.~\ref{subsec:ASproof}, let us introduce
a Dirac operator on a fermion field in $X\times \mathbb{R}$,
\begin{align}
D_{\rm DW}^{2n+1}(M)&=\tau_2 \otimes\partial_{s} + i\tau_1 \otimes \gamma_5(D+m(x,s))\nonumber\\
&=i\left(
\begin{array}{cc}
0 & \partial_{s}+\gamma_5(D+m(x,s))\\
-\partial_{s}+\gamma_5(D+m(x,s)) & 0
\end{array}
\right).
\end{align}
We may call it a massless $2n+2$-dimensional Dirac operator,
whose $2n+2$-th momentum is fixed to $m(x,s)$.
In either interpretation, this Dirac operator $D_{\rm DW}^{2n+1}(M)$ has
a chiral symmetry with $\gamma_7 = \tau_3 \otimes 1_{4\times 4}$.

Now let us solve the Dirac equation,
\begin{align}
  \label{eq:DDWeq}
D_{\rm DW}^{2n+1}(M)\phi(x,s)&=\left[\tau_2 \otimes\partial_{s} + i\tau_1 \otimes \gamma_5(D+m(x,s))\right]\phi(x,s)=0.
\end{align}
In the large $M$ limit, the  zero mode solution must shrink
around the domain-wall.
For $x \in X_+$ and $s\sim 0$, they satisfy
\begin{align}
  \label{eq:DDWsol}
\tau_2 [1\otimes\partial_{s} - \tau_3 \otimes \gamma_5 M{\rm sgn}(s)]\phi(x,s)=0,
\end{align}
to which we find a edge-localized solution \cite{Jackiw:1975fn, Callan:1984sa}
satisfying
\begin{align}
\phi(x,s)\propto \exp(-M|s|),\;\;\;-\tau_3\otimes\gamma_5 \phi(x,s)=+ \phi(x,s)
\end{align}
and the massless Dirac equation $D\phi(x,s)=0$.
Note that $\gamma_7 = -\tau_3 \otimes 1_{4\times 4}$ and
$1_{2\times 2}\otimes \gamma_5$
must have the same eigenvalue.

For $x$ near the original domain-wall between $X_+$ and $X_-$,
and $s>0$, the zero modes must satisfy
\begin{align}
i\tau_1 \otimes \gamma_5\gamma_4[\partial_{x_4} -\gamma_4 M{\rm sgn}(x_4)]\phi(x,s)=0,
\end{align}
where we denote the normal direction to the domain-wall by $x_4$,
and the corresponding gamma matrix by $\gamma_4$.
To this equation, we have edge-localized solution
\begin{align}
  \label{eq:DDWsol2}
\phi(x,s)\propto \exp(-M|x_4|), 
\end{align}
which is chiral, $-1\otimes\gamma_4 \phi(x,s)=+ \phi(x,s)$.
The zero modes also satisfy
\begin{align}
&\tau_2\otimes 1 \left[1\otimes 1 \partial_{s} + \tau_3 \otimes \gamma_5(1\otimes \gamma_iD^i)\right]\phi(x,s)\nonumber\\
  &=\tau_2\otimes 1 \left[1\otimes 1 \partial_{s} - (1\otimes 
      \textcolor{black}{\gamma_4}\gamma_iD^i)
    (-\tau_3 \otimes \gamma_5)
    \right]\phi(x,s)=0.
\end{align}
Since $\gamma_4\gamma_iD^i$ is $s$-independent, it is natural to assume that
the requirement $-\tau_3 \otimes \gamma_5\phi(x,s=0) =\phi(x,s=0)$ at $s=0$ is
inherited to the all range of $s>0$.

Here let us define $A=\gamma_4\gamma_iD^i$ and its eigenvalues and states
by $\lambda_A$ and $\phi_{\lambda_A}(\bm{x})$, respectively.
The above Dirac equation is now simplified as
\begin{align}
(\partial_{s} -  A) \phi(x)=0,
\end{align}
and the solution is given by a linear combination of $\phi_{\lambda_A}(\bm{x})$,
\begin{align}
\phi(x)= \sum_{\lambda_A} \alpha_{\lambda_A} \phi_{\lambda_A}(\bm{x})\exp(\lambda_A s).
\end{align}
Note that the coefficient $\alpha_{\lambda_A}$ must be zero for   $\lambda_A>0$,
otherwise, the solution is not normalizable.
Interestingly, this constraint is exactly the same as
the APS boundary condition at the boundary of $X_+$ and $s=0$.

This is not a coincidence but proved in
the original APS paper\cite{Atiyah:1975jf,Atiyah:1976jg,Atiyah:1980jh}:
the APS index is equal to the index on a manifold
with an infinite cylinder attached to the original boundary,
with respect to the square integrable modes.
Here, the gauge field and metric are just copied along the cylinder\footnote{
  To be precise, we have to take the induced metric embedded at the ``corner''
  $s=0$ and boundary of $X_+$ into account. In our paper~\citen{Fukaya:2019qlf},
  it was proved that the difference of the  metric from the original APS
  is small enough to guarantee that the index is unchanged.
}.

To summarize, the solution to $D_{\rm DW}^{2n+1}(M)\phi(x,s)=0$ with
chirality $\gamma_7=\pm 1$ must satisfy $D\phi(x,s=0)=0$
with the same chirality of $\gamma_5$
in the region $x\in X_+$ and satisfy the APS boundary condition at
the boundary. Namely, we have shown that
${\rm Ind}(D_{\rm DW}^{2n+1}(M)) ={\rm Ind}_{\rm APS}(D|_{X_+})$.

Next we consider the same Dirac equation but at $s=\pm\infty$.
At $s=-\infty$, the equation is
\begin{align}
  0&=\left[\tau_2 \otimes\partial_{x_5} + i\tau_1 \otimes \gamma_5(D+M)\right]\phi(x)
  = \tau_2 \left[1 \otimes\partial_{x_5} + \tau_3 \otimes H_{\rm PV}\right]\phi(x).
\end{align}
For $\gamma_7 =\tau_3=\pm 1$, the solution is a linear combination of
\[
\exp(\mp \lambda_{\rm PV}s),
\]
where $\lambda_{\rm PV}$ is an eigenvalue of $H_{\rm PV}$, which is
normalizable only when ${\rm sgn}(\lambda_{\rm PV}) = \mp 1$,
in the $s\to -\infty$ limit.
At $s=+\infty$, the domain-wall fermion Dirac operator appears,
\begin{align}
  0&=\left[\tau_2 \otimes\partial_{x_5} + i\tau_1 \otimes
    \gamma_5(D-\varepsilon M)\right]\phi(x)
  = \tau_2 \left[1 \otimes\partial_{x_5} + \tau_3 \otimes H_{\rm DW}\right]\phi(x),
\end{align}
where $\varepsilon(x)=+1$ for $x\in X_+$ and -1, otherwise.
For $\gamma_7 =\tau_3=\pm 1$, the solution is a linear combination of
\[
\exp(\mp \lambda_{\rm DW}s),
\]
where $\lambda_{\rm DW}$ is an eigenvalue of $H_{\rm DW}$, which is
normalizable only when ${\rm sgn}(\lambda_{\rm DW}) = \pm 1$,
in the $s\to \infty$ limit.

Now the proof goes the exactly same way as in the AS index
shown in Sec.~\ref{subsec:ASproof}.
Smoothing the step function
and considering an adiabatic
$s$ dependence of the solutions, we can
assign a one-to-one correspondence between the solution
and the one-parameter family of each eigenvalue $\lambda_s$
of the Hermitian Dirac operator on the slice $s$.
For a $\gamma_7=+1$ zero mode,
we have to find a $\lambda_s$ with $\lambda_{\rm PV}<0$ (at $s=-\infty$)
and $\lambda_{\rm DW}>0$ (at $s=+\infty$), while for $\gamma_7=-1$
we have to find the one with $\lambda_{\rm PV}>0$ and $\lambda_{\rm DW}<0$.
The index is given by the spectral flow along the path $s\in [-\infty,+\infty]$
counting the increased number of the positive eigenvalues
subtracted by that of the negative eigenvalues divided by 2:
\begin{align}
  {\rm Ind}[D_{\rm DW}^{2n+1}(M)]&=
  \frac{1}{2}\sum_{\lambda_{\rm PV}}{\rm sgn}(\lambda_{\rm PV})-\frac{1}{2}\sum_{\lambda_{\rm DW}}{\rm sgn}(\lambda_{\rm DW}),
\end{align}
which is equal to the RHS of Eq.~(\ref{eq:APSeta}).
The proof for our main theorem is complete.

In this section, have perturbatively and non-perturbatively 
shown that
\begin{align}
  \left({\rm Ind}(D_{\rm DW}^{2n+1}(M)) =\right) {\rm Ind}_{\rm APS}(D|_{X_+}) = -\frac{1}{2}\eta(H_{\rm DW})+\frac{1}{2}\eta(H_{\rm PV}).
\end{align}  
For the massive expression, the chiral symmetry is not crucial at all
and we do not need any nonlocal boundary conditions.
In the next section, we will see that this massive
formulation is valid even on a lattice.

\section{APS index on a lattice}
\label{sec:APSlattice}

After a long history of chiral symmetry in lattice gauge theory
since Refs.~\citen{Nielsen:1980rz,Nielsen:1981xu},
it is now well-known that the Atiyah-Singer index on an even-dimensional
torus can be formulated even
with finite lattice spacings\cite{Hasenfratz:1998ri}.
This is possible using the overlap Dirac operator\cite{Neuberger:1997fp}
or that in the perfect action\cite{Hasenfratz:1993sp},
satisfying the Ginsparg-Wilson relation\cite{Ginsparg:1981bj},
as a consequence of the exactness of the modified
lattice chiral symmetry\cite{Luscher:1998pqa}.
However, its extension to the APS index has been untouched
due to the difficulty of the nonlocal boundary condition.
In this section, we show that the domain-wall Dirac operator
can formulate the APS index even on a lattice\cite{Fukaya:2019myi}.

The difficulty of the chiral symmetry originates
from the periodic boundary condition of the
momentum space in lattice gauge theory.
The naive Dirac operator $i\gamma^\mu p_\mu$ is replaced by
$i\gamma^\mu \sin (p_\mu a)/a$ with a lattice spacing $a$,
which has 15 unphysical poles,
as $p_\mu$ for each direction $\mu$ has two zero points $p_\mu=0$
and $p_\mu=\pi/a$. To remove the unwanted modes called
doublers, one needs to add a momentum-dependent
mass term (Wilson term), which explicitly breaks
the chiral symmetry.
Since the Nielsen-Ninomiya theorem\cite{Nielsen:1980rz,Nielsen:1981xu}
proved that the chiral symmetry must be broken
in order to remove the doublers, it has been a challenge
to formulate an exactly chiral symmetric fermion on a lattice.

In fact, there is one symmetry 
which must be explicitly broken
on a lattice. It is the axial $U(1)$ symmetry.
If there is a lattice Dirac operator free from doublers
which breaks the axial $U(1)$ only,
we are able to respect all other part of chiral symmetries
$SU(N_f)_L \times SU(N_f)_R \times U(1)_V$,
where $N_f$ is the number of flavors, in a lattice gauge theory.

The overlap Dirac operator\cite{Neuberger:1997fp}
perfectly meets the desired conditions.
It is defined by
\begin{align}
D_{ov} = \frac{1}{a}\left(1+\gamma_5\frac{H_W}{\sqrt{H_W^2}}\right),
\end{align}  
where $H_W = \gamma_5(D_W-M)$ is a massive Wilson Dirac operator
with a cut-off scale mass $M=1/a$, and satisfies
the Ginsparg-Wilson relation
\begin{align}
  \gamma_5 D_{ov} + D_{ov} \gamma_5 = a D_{ov}\gamma_5 D_{ov}.
\end{align}
The overlap fermion action
$S=\sum_x \bar{\psi}D_{ov}\psi(x)$
is invariant under the modified chiral transformation,
\begin{align}
  \psi\to e^{i\alpha \gamma_5(1-aD_{ov})}\psi,\;\;\;\bar{\psi}\to\bar{\psi}e^{i\alpha\gamma_5},
\end{align}
but the fermion measure transforms as
\begin{align}
  d\psi d\bar{\psi} \to
  \exp\left[i\alpha{\rm Tr}(\gamma_5+\gamma_5(1-aD_{ov}))\right]
  d\psi d\bar{\psi},
\end{align}
which precisely reproduces the axial $U(1)$ anomaly\footnote{
  For the flavor-non-singlet transformation, the measure is invariant.
  Namely, the $SU(N_f)_L \times SU(N_f)_R $ is still preserved.
}.
Moreover,
\begin{align}
  \label{eq:latAS}
{\rm Tr}\gamma_5\left(1-\frac{aD_{ov}}{2}\right),
\end{align}
corresponds to the Atiyah-Singer index.
It is known that the eigenvalue spectrum of the overlap Dirac
operator makes a circle of radius $1/a$
whose center is located at $1/a$ in the complex plane.
We can show that the complex eigenvalues of $D_{\rm ov}$
make $\pm$ pairs of the ``chirality'' operator
$\gamma_5\left(1-\frac{aD}{2}\right)$ and the real modes
at $2/a$, which corresponds to the doublers, do not contribute.
Thus the above trace essentially counts the index of $D_{ov}$.

Despite a remarkable success of the AS index in lattice gauge theory,
its extension to the APS has been not known.
In continuum theory, the Dirac operator can be decomposed
as $D=\gamma^4D_4 +\gamma^iD_i$, and the APS boundary condition
is imposed with respect to the eigenfunctions of $\gamma_4\gamma^iD_i$,
which corresponds to the Dirac operator on the boundary.
For the lattice Dirac operator $D_{ov}$,
there is no simple way to separate the boundary part of the operator.
Moreover, it was shown in Ref.~\citen{Luscher:2006df} that
any boundary condition but periodic or anti-periodic boundary conditions
breaks the Ginsparg-Wilson relation.

Although formulating the APS index of a massless
lattice Dirac operator looks hopeless,
a great hint is hidden in the formulation of the AS index.
Simply substituting the definition of the overlap Dirac operator
into Eq.~(\ref{eq:latAS}), we end up with
the $\eta$ invariant of the massive Wilson Dirac operator\footnote{
  This relation was known in Refs.~\citen{Itoh:1987iy, Adams:1998eg}.
  However, as far as we know, its mathematical importance of using
  the massive Dirac operator
  without chiral symmetry has not been discussed.
  See Refs.~\citen{Yamashita:2020nkf,Kubota:2020tpr}
  for recent progress by mathematicians. 
  },
\begin{align}
  {\rm Ind}(D_{ov})=-\frac{1}{2}{\rm Tr}\frac{H_W}{\sqrt{H_W^2}}
  = -\frac{1}{2}\eta(\gamma_5(D_W-M)).
\end{align}
Surprisingly, the lattice AS index theorem ``knew'' that
the index can be given with massive Dirac operator
and chiral symmetry is not important: the Wilson Dirac
operator is enough to define it.

Let us here summarize the situation in
Tab.~\ref{tab:massless} and \ref{tab:massive}.
The standard index theorems with respect to the massless fermion Dirac operator
requires much efforts to formulate.
The APS index elaborates a nonlocal (and unphysical) boundary condition,
the lattice AS needs a complicated chiral Dirac operator
satisfying the Ginsparg-Wilson relation,
and the lattice version of APS is not even known yet.
Instead, the $\eta$ invariant of the massive Dirac operator offers 
a unified description of the index theorems on a closed manifold.
The APS index is obtained by just changing the sign of
the mass on a domain-wall.
The lattice version of the AS index is given by
the Wilson Dirac operator,
even without knowing its equivalence to the index of the overlap Dirac operator.
As already shown in Fig.~\ref{fig:ASspecflow2},
the massive bulk definition is stable against symmetry breaking
of the original Dirac operator, as it counts 
one-dimensional objects, the spectral flow, rather
than points of the massless Dirac operator spectrum.
It is now natural to speculate that the lattice formulation of the APS
index is given by the $\eta$ invariant of lattice domain-wall Dirac operator
\begin{align}
  -\frac{1}{2}\eta(\gamma_5(D_W-\varepsilon M)).
\end{align}

In Ref.~\citen{Fukaya:2019myi}, by a tedious but straight-forward perturbative computation\footnote{
  Bulk part computation is similar to that of AS index \cite{Suzuki:1998yz}.
  },
we have shown on a four-dimensional
Euclidean lattice with periodic boundary conditions (of which continuum limit is $T^4$)
that
\begin{align}
-\frac{1}{2}\eta(\gamma_5(D_W-\varepsilon M_1+M_2)) =& \frac{1}{32\pi^2}\int_{\textcolor{black}{0<x_4<T}} d^4x 
\epsilon_{\mu\nu\rho\sigma}{\rm tr}_cF^{\mu\nu}F^{\rho\sigma}(x)
\nonumber\\
&+\frac{1}{2}\eta(iD^{\rm 3D})|_{x_4=0}-\frac{1}{2}\eta(iD^{\rm 3D})|_{x_4=T}+O(a),
\end{align}
where $\varepsilon={\rm sgn}(x_4-a/2){\rm sgn}(T-x_4-a/2)$ represent two domain-walls we put
at $x_4=a/2$ and $x_4=T-a/2$ with the lattice spacing $a$.
The LHS is shown in the limit $M_1+M_2\to \infty$ while $M=M_1-M_2$ is fixed.
Note that the left-hand side is always an integer.
Therefore, we expect that this definition of the APS index on the lattice is
nonperturbatively valid, with possibly additional requirement on the
smoothness of the gauge link variables\cite{Luscher:1998kn}.

\begin{table}[tbh]
  \tbl{The standard formulation with massless Dirac operator}{
  \centering
  \begin{tabular}{|c|c|c|}
    \hline
    & continuum & lattice\\\hline
    AS & ${\rm Tr}\gamma_5e^{D^2/M^2}$ & ${\rm Tr}\gamma_5(1-aD_{ov}/2)$\\\hline
    APS  & ${\rm Tr}\gamma_5e^{D^2/M^2}$ w/ APS b.c. & not known.\\\hline
  \end{tabular}
    \label{tab:massless}}
\end{table}
  \begin{table}[tbh]   
  \tbl{The $\eta$ invariant of massive Dirac operator}{
  \centering
  \begin{tabular}{|c|c|c|}
    \hline
    & continuum & lattice\\\hline
    AS & $-\frac{1}{2}\eta(\gamma_5(D- M))^{\rm PV.}$ & $-\frac{1}{2}\eta(\gamma_5(D_W-M))$\\\hline
    APS  & $-\frac{1}{2}\eta(\gamma_5(D-\varepsilon M))^{\rm PV.}$ & $-\frac{1}{2}\eta(\gamma_5(D_W-\varepsilon M))$ \\\hline
  \end{tabular}
  \label{tab:massive}}
\end{table}

\section{Mod-two APS index}
\label{sec:APSmod2}

So far we have only considered the standard index taking
an integer value on an
even-dimensional manifold.
In Ref.~\citen{Fukaya:2020tjk}, we extended our work to
the odd-dimensional cases, which have the so-called mod-two index
defined by the number of zero modes mod 2.
The mod-two index makes sense only when the Dirac operator
is real where every nonzero mode makes a $\pm$ pair:
for $D\phi=i\lambda\phi$, we have $D\phi^*=-i\lambda\phi^*$.
If one non-zero mode goes down to zero,
so does its pair with the opposite sign.
The number of zero modes can change by even numbers only
under smooth variation of the gauge field and metric.

In the odd-dimensional case, we cannot use the $\eta$ invariant.
Instead, we have the so-called spectral flow.
As we have seen in Sec.~\ref{sec:APSDW}, the $\eta$ invariant
is also equivalent to the spectral flow\footnote{
This is true only in even dimensions.
}:
\begin{align}
  {\rm Sf}[H_1]_{H_0} = -\frac{1}{2}\eta(H_1)+\frac{1}{2}\eta(H_0),
\end{align}
where the two Hermitian operator $H_0$ and  $H_1$ are
smoothly connected by a one-parameter family $H_t$.

For a real Dirac operator $D$ (in odd dimensions),
we introduce a doubled spinor field and an anti-Hermitian and real
massive Dirac operator on it,
\begin{align}
A(m):=\left(
\begin{array}{cc}
&D+m\\
-(D+m)^\dagger &
\end{array}
\right).
\end{align}
and using the mod-two spectral flow\cite{Carey,Carey20},
${\rm Sf}_{\rm 2}[A_1]_{A_0}$, which is defined by
the number of zero-crossing pairs along  the one-parameter family $A_t$ $t\in[0,1]$,
we have proved that
\begin{align}
{\rm Sf}_{\rm 2}[A(-\varepsilon M)]_{A(M)},
\end{align}
is equal to the mod-two APS index ${\rm Ind}^{\rm mod-2}_{\rm APS}(D)$ in the $\varepsilon=+1$ region.

Moreover, we showed that the index is
related to the domain-wall fermion determinant by
\begin{align}
\det \frac{D-\varepsilon M}{D+M}\propto (-1)^{{\rm Ind}^{\rm mod-2}_{\rm APS}(D)},
\end{align}
which allows us to decompose the edge and bulk contributions as
\begin{eqnarray}
  {\rm Ind}_{\rm APS}(D) &=& I_{\rm edge} + I_{\rm bulk}\;\;\;(\mbox{mod 2}),\nonumber\\
  I_{\rm edge} &=& \frac{1-{\rm sgn}\left[\det D_{\rm edge}\right]}{2},\nonumber\\
  I_{\rm bulk} &=& \frac{1-{\rm sgn}\left[\det D_{\rm bulk}\right]}{2}.
\end{eqnarray}
Here,
\begin{align}
D_{\rm edge} &:= \left(
    \begin{array}{cc}
      D- \varepsilon M & 0\\
      0 & \partial+ \varepsilon M
      \end{array}\right)\left(
    \begin{array}{cc}
      D- \varepsilon M & \mu I\\
      \mu I^{-1}& \partial+ \varepsilon M
    \end{array}\right)^{-1},\\
 D_{\rm bulk}&:=\left(
    \begin{array}{cc}
      D- \varepsilon M & \mu I\\
      \mu I^{-1}& \partial+ \varepsilon M
    \end{array}
         \right)\left(
     \begin{array}{cc}
       D+M & 0 \\
       0 & \partial+M
     \end{array}
     \right)^{-1},  
\end{align}
where $\partial$ denotes the free Dirac operator
and the mass $\mu\ll M$, which breaks the gauge symmetry,
is introduced to give an ultra-violet cut-off to the
edge modes, while it is the infra-red cut-off
for the bulk modes.
Since $D_{\rm edge}$ and $D_{\rm bulk}$ are both real,
the decomposition represents the inflow of
the global anomaly\cite{Witten:1982fp,Lott:1988mt,Witten:2016cio,Wang:2018qoy}.

Note that the lattice Wilson Dirac operator inherits
the real structure of the operator, and
the lattice version of $A(m)$,
\begin{align}
A_W(m):=\left(
\begin{array}{cc}
&D_W+m\\
-(D_W+m)^\dagger &
\end{array}
\right)
\end{align}
and the corresponding mod-two spectral flow
from the spectrum of a reference operator $A_W(M)$
can be defined without any difficulty.
Therefore, as shown in Tab.~\ref{tab:massive2},
this achieves a further unification of the index theorems:
the standard/mod-two AS/APS index in continuum/lattice theory can be
reformulated  by the spectral flow of a massive Dirac operator
on a closed manifold without boundary.

  \begin{table}[tbh]   
  \tbl{The spectral flow of massive Dirac operator}{
  \centering
  \begin{tabular}{|c|c|c|}
    \hline
    & continuum & lattice\\\hline
    AS & ${\rm Sf}[\gamma_5(D- M)]_{H(M)}$ & ${\rm Sf}[\gamma_5(D_W-M)]_{H_W(M)}$\\\hline
    APS  & ${\rm Sf}[\gamma_5(D-\varepsilon M)]_{H(M)}$ & ${\rm Sf}(\gamma_5(D_W-\varepsilon M)]_{H_W(M)}$ \\\hline
    mod-2 AS & ${\rm Sf}_2\left[\left(
\begin{array}{cc}
&D-M\\
-(D-M)^\dagger &
\end{array}
\right)\right]_{A(M)}$ &
    ${\rm Sf}_2\left[\left(
\begin{array}{cc}
&D_W-M\\
-(D_W-M)^\dagger &
\end{array}
\right)\right]_{A_W(M)}
$\\\hline
    mod-2 APS  & ${\rm Sf}_2\left[\left(
\begin{array}{cc}
&D-\varepsilon M\\
-(D-\varepsilon M)^\dagger &
\end{array}
\right)\right]_{A(M)}$ & ${\rm Sf}_2\left[\left(
\begin{array}{cc}
&D_W-\varepsilon M\\
-(D_W-\varepsilon M)^\dagger &
\end{array}
\right)\right]_{A_W(M)}
$
\\\hline
  \end{tabular}
  \label{tab:massive2}}
\end{table}

\section{Summary and discussion}
\label{sec:summary}

In this review, we have shown that the axial anomaly and index theorems
can be understood with massive fermions.
The $\eta$ invariant or spectral flow of the massive Dirac operator,
including the domain-wall Dirac operator,
on a closed manifold without boundary,
gives a unified view of different types of index theorems.
The formulation is physicist-friendly enough
that the application to lattice gauge theory is straightforward.

The massive formulation of the index theorems
have a tight connection to physics of topological insulators.
The edge-localized modes appear at the domain-wall,
and describe the $\eta$ invariant
of the boundary operator.
Unlike the original formulation of APS,
the non-locality of the $\eta$ invariant is
not a consequence of the boundary condition, but
reflects the massless nature of the edge modes.
From the bulk mode contribution, which is massive,
we obtain a local expression of the curvature term.

In order to identify the topological and normal
phases of massive fermions, it is essential
to choose a regularization of the $\eta$ invariant
that distinguishes the sign of the mass.
In this paper, we have chosen the Pauli-Villars subtraction
in continuum theory, and Wilson Dirac operator on a lattice.
We have set the coefficients of the Pauli-Villars mass and the Wilson term
positive, and the domain of the topological phase
is identified as that having negative fermion mass.
It may be difficult to formulate the anomaly (inflow)
with regularizations which cannot distinguish the sign of the mass,
such as dimensional or heat kernel regularizations.

The equivalence of the massive formulation to the
original AS or APS has been proven through the index
of a higher-dimensional Dirac operator $D_{\rm DW}^{2n+1}(M)$ on $X\times \mathbb{R}$.
Since the extended manifold has cylindrical ends at
$s\to \pm\infty$, the index is equal to
the index on $X\times [-1,1]$, where the APS boundary condition
is imposed at $s=\pm 1$. Then an interesting question is if
we can express this higher-dimensional index by
the $\eta$ invariant again:
\begin{equation}
  {\rm Ind}_{\rm APS}(D_{\rm DW}^{2n+1}(M)|_{X\times [-1,1]})
  = -\frac{1}{2}\eta(\gamma_7(D_{\rm DW}^{2n+1}(M)-\mu\kappa))^{\rm PV.},
\end{equation}
where we have introduced a second mass term $\mu\kappa$
with $\kappa=1$ in $s \in [-1,1]$, and $-1$, otherwise.
In this ``doubly'' domain-wall fermion,
the original edge mode localized in $2n-1$ dimensional manifold $Y$ becomes
the edge-of-edge states\cite{Neuberger:2003yg,Fukaya:2016ofi,Hashimoto:2017tuh}
of $\gamma_7(D_{\rm DW}^{2n+1}(M)-\mu\kappa)$,
which is localized at the junction of the first and second domain-walls.
This recursive structure might be useful for understanding
physics of higher order topological insulators\cite{hoti1,Schindler:2017etn}.

Our work is limited to the real and pseudo-real fermions.
In the literature\cite{Dai:1994kq,Yonekura:2016wuc,Witten:2019bou}, it is claimed that
the phase of the massive complex fermion is
controlled by the $\eta$ invariant
of a massless Dirac operator with the APS boundary condition.
It would be interesting to study if the domain-wall fermion
can reformulate the phase of the complex fermion determinant
without any nonlocal boundary conditions.

As a final discussion, let us reconsider the difference between
massless and massive systems.
In physics, we have been imprinted that a symmetry
makes some particle fields massless, which leads to
a good control of understanding nature.
For example, the gauge symmetry in the Yang-Mills
fields makes gauge bosons massless, which
guarantees that the theory renormalizable.
Even when the symmetry is broken, if it is spontaneous,
we still have a good control of the theory,
thanks to the Nambu-Goldstone theorem\cite{Nambu:1960tm,Goldstone:1961eq}.
Such a ``symmetry fundamentalism'' that
massless is better than massive, has been
a driving force in particle physics and successful for 50 years.
However, we have also experienced problems where
we cannot find any relevant symmetry to explain the small scales
of the particles, such as the Higgs boson mass.
In this review, we have seen that
massive fermion is not always inferior to the massless chiral fermion,
but sometimes gives mathematically equal or better understanding of physics.
In the systems we used in this article,
the symmetry appears  not as a guiding principle of the theory
but as an accidental consequence of the localization
of edge modes in the massive system.
It would be nice if our works could give a hint for
the ``post-symmetry'' era of new physics.

\section*{Acknowledgments}
The author thanks M.~Furuta, N.~Kawai, S.~Matsuo, Y.~Matsuki, M.~Mori, K.~Nakayama,
T.~Onogi, S.~Yamaguchi and M.~Yamashita for the exciting collaborations.
The author also thanks S.~Sugimoto and members of YITP, Kyoto Univ.
for their kind hospitality during his stay for the intensive lecture.
This work was supported in part by JSPS KAKENHI Grant Number
JP18H01216 and JP18H04484.



\begin{thebibliography}{0}    

\bibitem{Atiyah:1963zz}
  M.~F.~Atiyah and I.~M.~Singer,
  ``The index of elliptic operators on compact manifolds,''
  Bull.\ Am.\ Math.\ Soc.\  {\bf 69}, 422 (1963).
  doi:10.1090/S0002-9904-1963-10957-X

\bibitem{Atiyah:1968mp}
  M.~F.~Atiyah and I.~M.~Singer,
  ``The Index of elliptic operators. 1,''
  Annals Math.\  {\bf 87}, 484 (1968).
  doi:10.2307/1970715


\bibitem{Atiyah:1975jf}
  M.~F.~Atiyah, V.~K.~Patodi and I.~M.~Singer,
  ``Spectral asymmetry and Riemannian Geometry I,''
  Math.\ Proc.\ Cambridge Phil.\ Soc.\  {\bf 77}, 43 (1975).
  doi:10.1017/S0305004100049410

\bibitem{Atiyah:1976jg} 
  M.~F.~Atiyah, V.~K.~Patodi and I.~M.~Singer,
  ``Spectral asymmetry and Riemannian geometry II,''
  Math.\ Proc.\ Cambridge Phil.\ Soc.\  {\bf 78}, 405 (1975).
  doi:10.1017/S0305004100051872

\bibitem{Atiyah:1980jh} 
  M.~F.~Atiyah, V.~K.~Patodi and I.~M.~Singer,
  ``Spectral asymmetry and Riemannian geometry. III,''
  Math.\ Proc.\ Cambridge Phil.\ Soc.\  {\bf 79}, 71 (1976).
  doi:10.1017/S0305004100052105


\bibitem{Adler:1969gk}
S.~L.~Adler,
``Axial vector vertex in spinor electrodynamics,''
Phys. Rev. \textbf{177}, 2426-2438 (1969)
doi:10.1103/PhysRev.177.2426
\bibitem{Bell:1969ts}
J.~Bell and R.~Jackiw,
``A PCAC puzzle: $\pi^0 \to \gamma \gamma$ in the $\sigma$ model,''
Nuovo Cim. A \textbf{60}, 47-61 (1969)
doi:10.1007/BF02823296


  


\bibitem{Jackiw:1975fn}
R.~Jackiw and C.~Rebbi,
``Solitons with Fermion Number 1/2,''
Phys. Rev. D \textbf{13}, 3398-3409 (1976)
doi:10.1103/PhysRevD.13.3398

\bibitem{Callan:1984sa}
C.~G.~Callan, Jr. and J.~A.~Harvey,
``Anomalies and Fermion Zero Modes on Strings and Domain Walls,''
Nucl. Phys. B \textbf{250}, 427-436 (1985)
doi:10.1016/0550-3213(85)90489-4

\bibitem{Kaplan:1992bt} 
  D.~B.~Kaplan,
  ``A Method for simulating chiral fermions on the lattice,''
  Phys.\ Lett.\ B {\bf 288}, 342 (1992)
  doi:10.1016/0370-2693(92)91112-M
  [hep-lat/9206013].



  

\bibitem{Fukaya:2017tsq}
H.~Fukaya, T.~Onogi and S.~Yamaguchi,
``Atiyah-Patodi-Singer index from the domain-wall fermion Dirac operator,''
Phys. Rev. D \textbf{96}, no.12, 125004 (2017)
doi:10.1103/PhysRevD.96.125004
[arXiv:1710.03379 [hep-th]];
``Atiyah-Patodi-Singer index theorem for domain-wall fermion Dirac operator,''
EPJ Web Conf. \textbf{175}, 11009 (2018)
doi:10.1051/epjconf/201817511009
[arXiv:1712.03679 [hep-lat]].

\bibitem{Fukaya:2019qlf}
H.~Fukaya, M.~Furuta, S.~Matsuo, T.~Onogi, S.~Yamaguchi and M.~Yamashita,
``The Atiyah-Patodi-Singer index and domain-wall fermion Dirac operators,''
Commun. Math. Phys. \textbf{380}, no.3, 1295-1311
doi:10.1007/s00220-020-03806-0
[arXiv:1910.01987 [math.DG]].



\bibitem{Fukaya:2019myi}
H.~Fukaya, N.~Kawai, Y.~Matsuki, M.~Mori, K.~Nakayama, T.~Onogi and S.~Yamaguchi,
``Atiyah-Patodi-Singer index on a lattice,''
PTEP \textbf{2020}, no.4, 043B04 (2020)
doi:10.1093/ptep/ptaa031
[arXiv:1910.09675 [hep-lat]].

\bibitem{Onogi:2021slv}
T.~Onogi and T.~Yoda,
``Comments on the Atiyah-Patodi-Singer index theorem, domain wall, and Berry phase,''
JHEP \textbf{12}, 096 (2021)
doi:10.1007/JHEP12(2021)096
[arXiv:2109.08274 [hep-th]].


\bibitem{Fukaya:2020tjk}
H.~Fukaya, M.~Furuta, Y.~Matsuki, S.~Matsuo, T.~Onogi, S.~Yamaguchi and M.~Yamashita,
``Mod-two APS index and domain-wall fermion,''
[arXiv:2012.03543 [hep-th]].



\bibitem{Witten:2015aba}
E.~Witten,
``Fermion Path Integrals And Topological Phases,''
Rev. Mod. Phys. \textbf{88}, no.3, 035001 (2016)
doi:10.1103/RevModPhys.88.035001
[arXiv:1508.04715 [cond-mat.mes-hall]].

\bibitem{tHooft:1979rat}
G.~'t Hooft,
``Naturalness, chiral symmetry, and spontaneous chiral symmetry breaking,''
NATO Sci. Ser. B \textbf{59}, 135-157 (1980)


\bibitem{Hatsugai}
Y. Hatsugai, ``Chern number and edge states in the integer quantum Hall effect,'' Phys. Rev. Lett. 71, no. 22, 3697 (1993). doi:10.1103/PhysRevLett.71.3697; ``Edge states in the integer quantum Hall effect and the Riemann surface of the Bloch function,'' Phys. Rev. B 48, no.
16, 11851 (1993). doi:10.1103/PhysRevB.48.11851



\bibitem{Kurkov:2018pjw}
M.~Kurkov and D.~Vassilevich,
``Gravitational parity anomaly with and without boundaries,''
JHEP \textbf{03}, 072 (2018)
doi:10.1007/JHEP03(2018)072
[arXiv:1801.02049 [hep-th]].


\bibitem{Tachikawa:2018njr}
Y.~Tachikawa and K.~Yonekura,
``Why are fractional charges of orientifolds compatible with Dirac quantization?,''
SciPost Phys. \textbf{7}, no.5, 058 (2019)
doi:10.21468/SciPostPhys.7.5.058
[arXiv:1805.02772 [hep-th]].

\bibitem{Vassilevich:2018aqu}
D.~Vassilevich,
``Index Theorems and Domain Walls,''
JHEP \textbf{07}, 108 (2018)
doi:10.1007/JHEP07(2018)108
[arXiv:1805.09974 [hep-th]].

\bibitem{Garcia-Etxebarria:2018ajm}
I.~Garc\'\i{}a-Etxebarria and M.~Montero,
``Dai-Freed anomalies in particle physics,''
JHEP \textbf{08}, 003 (2019)
doi:10.1007/JHEP08(2019)003
[arXiv:1808.00009 [hep-th]].

\bibitem{Yonekura:2019vyz}
K.~Yonekura,
``Anomaly matching in QCD thermal phase transition,''
JHEP \textbf{05}, 062 (2019)
doi:10.1007/JHEP05(2019)062
[arXiv:1901.08188 [hep-th]].


\bibitem{Hsieh:2020jpj}
C.~T.~Hsieh, Y.~Tachikawa and K.~Yonekura,
``Anomaly inflow and $p$-form gauge theories,''
[arXiv:2003.11550 [hep-th]].


\bibitem{Hamada:2020mug}
Y.~Hamada, J.~M.~Pawlowski and M.~Yamada,
``Gravitational instantons and anomalous chiral symmetry breaking,''
[arXiv:2009.08728 [hep-th]].




\bibitem{Gromov:2015fda}
  A.~Gromov, K.~Jensen and A.~G.~Abanov,
  ``Boundary effective action for quantum Hall states,''
  Phys.\ Rev.\ Lett.\  {\bf 116}, no. 12, 126802 (2016)
  doi:10.1103/PhysRevLett.116.126802
  [arXiv:1506.07171 [cond-mat.str-el]].

\bibitem{Metlitski:2015yqa}
  M.~A.~Metlitski,
  ``$S$-duality of $u(1)$ gauge theory with $\theta =\pi$ on non-orientable manifolds: Applications to topological insulators and superconductors,''
  arXiv:1510.05663 [hep-th].

  
\bibitem{Seiberg:2016rsg}
N.~Seiberg and E.~Witten,
``Gapped Boundary Phases of Topological Insulators via Weak Coupling,''
PTEP \textbf{2016}, no.12, 12C101 (2016)
doi:10.1093/ptep/ptw083
[arXiv:1602.04251 [cond-mat.str-el]].

\bibitem{Tachikawa:2016xvs}
  Y.~Tachikawa and K.~Yonekura,
  ``Gauge interactions and topological phases of matter,''
  PTEP {\bf 2016}, no. 9, 093B07 (2016)
  doi:10.1093/ptep/ptw131
  [arXiv:1604.06184 [hep-th]].


\bibitem{Freed:2016rqq}
  D.~S.~Freed and M.~J.~Hopkins,
  ``Reflection positivity and invertible topological phases,''
  arXiv:1604.06527 [hep-th].

\bibitem{Yu:2017uqt}
  Y.~Yu, Y.~S.~Wu and X.~Xie,
  ``Bulk-edge correspondence, spectral flow and Atiyah-Patodi-Singer theorem for the Z2 -invariant in topological insulators,''
  Nucl.\ Phys.\ B {\bf 916}, 550 (2017)
  doi:10.1016/j.nuclphysb.2017.01.018
  [arXiv:1607.02345 [cond-mat.mes-hall]].


\bibitem{Hasebe:2016tjg}
  K.~Hasebe,
  ``Higher (Odd) Dimensional Quantum Hall Effect and Extended Dimensional Hierarchy,''
  Nucl.\ Phys.\ B {\bf 920}, 475 (2017)
  doi:10.1016/j.nuclphysb.2017.03.017
  [arXiv:1612.05853 [hep-th]].


\bibitem{Yonekura:2018ufj}
K.~Yonekura,
``On the cobordism classification of symmetry protected topological phases,''
Commun. Math. Phys. \textbf{368}, no.3, 1121-1173 (2019)
doi:10.1007/s00220-019-03439-y
[arXiv:1803.10796 [hep-th]].


\bibitem{Yao:2019ggu}
Y.~Yao and M.~Oshikawa,
``Generalized Boundary Condition Applied to Lieb-Schultz-Mattis-Type Ingappabilities and Many-Body Chern Numbers,''
Phys. Rev. X \textbf{10}, no.3, 031008 (2020)
doi:10.1103/PhysRevX.10.031008
[arXiv:1906.11662 [cond-mat.str-el]].


\bibitem{Fujikawa:2004cx}
K.~Fujikawa and H.~Suzuki,
``Path integrals and quantum anomalies,''
Oxford : Clarendon Press, 2004,
doi:10.1093/acprof:oso/9780198529132.001.0001

  
\bibitem{FlavourLatticeAveragingGroup:2019iem}
S.~Aoki \textit{et al.} [Flavour Lattice Averaging Group],
``FLAG Review 2019: Flavour Lattice Averaging Group (FLAG),''
Eur. Phys. J. C \textbf{80}, no.2, 113 (2020)
doi:10.1140/epjc/s10052-019-7354-7
[arXiv:1902.08191 [hep-lat]].










\bibitem{Witten:1982im}
E.~Witten,
``Supersymmetry and Morse theory,''
J. Diff. Geom. \textbf{17}, no.4, 661-692 (1982)

\bibitem{FurutaIndex}
Furuta, M.: Index theorem. 1, Translations of Mathematical Monographs, 235, Translated from the 1999
Japanese original by Kaoru Ono; Iwanami Series in Modern Mathematics, American Mathematical Society, Providence, RI, xviii+205, MR2361481 (2007)




\bibitem{Alvarez-Gaume:1984zst}
L.~Alvarez-Gaume, S.~Della Pietra and G.~W.~Moore,
``Anomalies and Odd Dimensions,''
Annals Phys. \textbf{163}, 288 (1985)
doi:10.1016/0003-4916(85)90383-5




\bibitem{Kobayashi:2021jbn}
S.~K.~Kobayashi and K.~Yonekura,
``Atiyah-Patodi-Singer index theorem from axial anomaly,''
doi:10.1093/ptep/ptab061
[arXiv:2103.10654 [hep-th]].

\bibitem{Luscher:2006df}
M.~Luscher,
``The Schrodinger functional in lattice QCD with exact chiral symmetry,''
JHEP \textbf{05}, 042 (2006)
doi:10.1088/1126-6708/2006/05/042
[arXiv:hep-lat/0603029 [hep-lat]].


\bibitem{Witten:2019bou}
E.~Witten and K.~Yonekura,
``Anomaly Inflow and the $\eta$-Invariant,''
[arXiv:1909.08775 [hep-th]].


\bibitem{Kanno:2021bze}
H.~Kanno and S.~Sugimoto,
``Anomaly and Superconnection,''
[arXiv:2106.01591 [hep-th]].




\bibitem{Nielsen:1980rz}
H.~B.~Nielsen and M.~Ninomiya,
``Absence of Neutrinos on a Lattice. 1. Proof by Homotopy Theory,''
Nucl. Phys. B \textbf{185}, 20 (1981)
[erratum: Nucl. Phys. B \textbf{195}, 541 (1982)]
doi:10.1016/0550-3213(82)90011-6

\bibitem{Nielsen:1981xu}
H.~B.~Nielsen and M.~Ninomiya,
``Absence of Neutrinos on a Lattice. 2. Intuitive Topological Proof,''
Nucl. Phys. B \textbf{193}, 173-194 (1981)
doi:10.1016/0550-3213(81)90524-1

\bibitem{Hasenfratz:1998ri}
P.~Hasenfratz, V.~Laliena and F.~Niedermayer,
``The Index theorem in QCD with a finite cutoff,''
Phys. Lett. B \textbf{427}, 125-131 (1998)
doi:10.1016/S0370-2693(98)00315-3
[arXiv:hep-lat/9801021 [hep-lat]].

\bibitem{Neuberger:1997fp}
H.~Neuberger,
``Exactly massless quarks on the lattice,''
Phys. Lett. B \textbf{417}, 141-144 (1998)
doi:10.1016/S0370-2693(97)01368-3
[arXiv:hep-lat/9707022 [hep-lat]].

\bibitem{Hasenfratz:1993sp}
P.~Hasenfratz and F.~Niedermayer,
``Perfect lattice action for asymptotically free theories,''
Nucl. Phys. B \textbf{414}, 785-814 (1994)
doi:10.1016/0550-3213(94)90261-5
[arXiv:hep-lat/9308004 [hep-lat]].

\bibitem{Ginsparg:1981bj} 
  P.~H.~Ginsparg and K.~G.~Wilson,
  ``A Remnant of Chiral Symmetry on the Lattice,''
  Phys.\ Rev.\ D {\bf 25}, 2649 (1982).
  doi:10.1103/PhysRevD.25.2649
  

\bibitem{Luscher:1998pqa} 
  M.~Luscher,
  ``Exact chiral symmetry on the lattice and the Ginsparg-Wilson relation,''
  Phys.\ Lett.\ B {\bf 428}, 342 (1998).
  doi:10.1016/S0370-2693(98)00423-7

\bibitem{Itoh:1987iy} 
  S.~Itoh, Y.~Iwasaki and T.~Yoshie,
  ``The U(1) Problem and Topological Excitations on a Lattice,''
  Phys.\ Rev.\ D {\bf 36}, 527 (1987).
  doi:10.1103/PhysRevD.36.527

\bibitem{Adams:1998eg} 
  D.~H.~Adams,
  ``Axial anomaly and topological charge in lattice gauge theory with overlap Dirac operator,''
  Annals Phys.\  {\bf 296}, 131 (2002)
  doi:10.1006/aphy.2001.6209.


\bibitem{Yamashita:2020nkf}
M.~Yamashita,
``A Lattice Version of the Atiyah\textendash{}Singer Index Theorem,''
Commun. Math. Phys. \textbf{385}, no.1, 495-520 (2021)
doi:10.1007/s00220-021-04021-1
[arXiv:2007.06239 [math.DG]].
  

\bibitem{Kubota:2020tpr}
Y.~Kubota,
``The index theorem of lattice Wilson--Dirac operators via higher index theory,''
[arXiv:2009.03570 [math-ph]].


  
\bibitem{Suzuki:1998yz}
H.~Suzuki,
``Simple evaluation of chiral Jacobian with overlap Dirac operator,''
Prog. Theor. Phys. \textbf{102}, 141-147 (1999)
doi:10.1143/PTP.102.141
[arXiv:hep-th/9812019 [hep-th]].


\bibitem{Luscher:1998kn}
M.~Luscher,
``Topology and the axial anomaly in Abelian lattice gauge theories,''
Nucl. Phys. B \textbf{538}, 515-529 (1999)
doi:10.1016/S0550-3213(98)00680-4
[arXiv:hep-lat/9808021 [hep-lat]].




\bibitem{Carey}
  A.~Carey, J.~Phillips and H.~Schulz-Baldes, ``{S}pectral flow for skew-adjoint {F}redholm operators,''
  J. Spectr. Theory 9 (2019), 137-170. doi: 10.4171/JST/243 [arXiv:1604.06994]

\bibitem{Carey20}
  C.~Bourne, A.~Carey, M.~Lesch and A.~Rennie,  ``The {KO}-valued spectral flow for skew-adjoint {F}redholm operators''
  Journal of Topology and Analysis (2020), 1--52. doi: 10.1142/S1793525320500557 	[arXiv:1907.04981 [math.KT]]


  
\bibitem{Witten:1982fp}
E.~Witten,
``An SU(2) Anomaly,''
Phys. Lett. B \textbf{117}, 324-328 (1982)
doi:10.1016/0370-2693(82)90728-6

\bibitem{Lott:1988mt}
J.~Lott,
``REAL ANOMALIES,''
J. Math. Phys. \textbf{29}, 1455-1464 (1988)
doi:10.1063/1.527940


\bibitem{Witten:2016cio}
E.~Witten,
``The "Parity" Anomaly On An Unorientable Manifold,''
Phys. Rev. B \textbf{94}, no.19, 195150 (2016)
doi:10.1103/PhysRevB.94.195150
[arXiv:1605.02391 [hep-th]].

\bibitem{Wang:2018qoy}
J.~Wang, X.~Wen and E.~Witten,
``A New SU(2) Anomaly,''
J. Math. Phys. \textbf{60}, no.5, 052301 (2019)
doi:10.1063/1.5082852
[arXiv:1810.00844 [hep-th]].
 


 

\bibitem{Neuberger:2003yg}
H.~Neuberger,
``Lattice chiral fermions from continuum defects,''
[arXiv:hep-lat/0303009 [hep-lat]].

\bibitem{Fukaya:2016ofi}
H.~Fukaya, T.~Onogi, S.~Yamamoto and R.~Yamamura,
``Six-dimensional regularization of chiral gauge theories,''
PTEP \textbf{2017}, no.3, 033B06 (2017)
doi:10.1093/ptep/ptx017
[arXiv:1607.06174 [hep-th]].

\bibitem{Hashimoto:2017tuh}
K.~Hashimoto, X.~Wu and T.~Kimura,
``Edge states at an intersection of edges of a topological material,''
Phys. Rev. B \textbf{95}, no.16, 165443 (2017)
doi:10.1103/PhysRevB.95.165443
[arXiv:1702.00624 [cond-mat.mes-hall]].

\bibitem{hoti1}
Wladimir A Benalcazar, B Andrei Bernevig, and Taylor L Hughes, “Quantized electric multipole insulators,”
Science \textbf{357}, 61--66 (2017), arXiv:1611.07987 [cond-mat].

\bibitem{Schindler:2017etn}
Frank Schindler, Ashley M Cook, Maia G Vergniory, Zhijun Wang, Stuart SP Parkin, B Andrei Bernevig, and
Titus Neupert, “Higher-order topological insulators,”
Science advances 4, eaat0346 (2018), arXiv:1708.03636
[cond-mat].



\bibitem{Dai:1994kq}
X.~Dai and D.~S.~Freed,
``eta invariants and determinant lines,''
J. Math. Phys. \textbf{35}, 5155-5194 (1994)
doi:10.1063/1.530747
[arXiv:hep-th/9405012 [hep-th]].

\bibitem{Yonekura:2016wuc}
K.~Yonekura,
``Dai-Freed theorem and topological phases of matter,''
JHEP \textbf{09}, 022 (2016)
doi:10.1007/JHEP09(2016)022
[arXiv:1607.01873 [hep-th]].



 
\bibitem{Nambu:1960tm}
Y.~Nambu,
``Quasiparticles and Gauge Invariance in the Theory of Superconductivity,''
Phys. Rev. \textbf{117}, 648-663 (1960)
doi:10.1103/PhysRev.117.648

\bibitem{Goldstone:1961eq}
J.~Goldstone,
``Field Theories with Superconductor Solutions,''
Nuovo Cim. \textbf{19}, 154-164 (1961)
doi:10.1007/BF02812722

\end{thebibliography}
\end{document}